\documentclass[sigconf]{acmart}
\renewcommand\footnotetextcopyrightpermission[1]{} % removes footnote with conference information in first column

\AtBeginDocument{%
  \providecommand\BibTeX{{%
    \normalfont B\kern-0.5em{\scshape i\kern-0.25em b}\kern-0.8em\TeX}}}

\usepackage{algorithm}
\usepackage[noend]{algorithmic}
\usepackage{multirow}
\usepackage{bm}
\usepackage{mathtools}
\usepackage{stmaryrd}
\usepackage{subfigure}

\newcommand{\commentout}[1]{}
\acmSubmissionID{290}

\begin{document}

\title{MISA: Online Defense of Trojaned Models using Misattributions}

\author{Panagiota Kiourti}
\email{pkiourti@bu.edu}
\affiliation{
  \institution{Boston University}
  \city{Boston}
  \state{MA}
  \country{USA}
}
\author{Wenchao Li}
\email{wenchao@bu.edu}
\affiliation{
  \institution{Boston University}
  \city{Boston}
  \state{MA}
  \country{USA}
}
\author{Anirban Roy}
\email{anirban.roy@sri.com}
\affiliation{
  \institution{SRI International}
  \city{Menlo Park}
  \state{CA}
  \country{USA}
}
\author{Karan Sikka}
\email{karan.sikka@sri.com}
\affiliation{
  \institution{SRI International}
  \city{Princeton}
  \state{NJ}
  \country{USA}
}
\author{Susmit Jha}
\email{susmit.jha@sri.com}
\affiliation{
  \institution{SRI International}
  \city{Menlo Park}
  \state{CA}
  \country{USA}
}

% \renewcommand{\shortauthors}{Kiourti, et al.}

\begin{abstract}
    Recent studies have shown that neural networks are vulnerable to Trojan attacks, where a network is trained to respond to specially crafted trigger patterns in the inputs in specific and potentially malicious ways. This paper proposes MISA, a new online approach to detect Trojan triggers for neural networks at inference time. Our approach is based on a novel notion called \textit{misattributions}, which captures the anomalous manifestation of a Trojan activation in the feature space. Given an input image and the corresponding output prediction, our algorithm first computes the model's attribution on different features.
It then statistically analyzes these attributions to ascertain the presence of a Trojan trigger. Across a set of benchmarks, we show that our method can effectively detect Trojan triggers for a wide variety of trigger patterns, including several recent ones for which there are no known defenses. Our method achieves 96\% AUC for detecting images that include a Trojan trigger without any assumptions on the trigger pattern.

% \commentout{This paper proposes a new approach to detecting neural Trojans on Deep Neural Networks during inference. This approach is based on monitoring the inference of a machine learning model, computing the attribution of the model's decision on different features of the input, and then statistically analyzing these attributions to detect whether an input sample contains the Trojan trigger. The anomalous attributions, aka misattributions, are then accompanied by reverse-engineering of the trigger to evaluate whether the input sample is truly poisoned with a Trojan trigger. We evaluate our approach on several benchmarks, including models trained on MNIST, Fashion MNIST, and German Traffic Sign Recognition Benchmark, and demonstrate the state of the art detection accuracy. 
% %We propose a method that monitors the decisions of a machine learning model, understands whether the input was Trojaned or not. We use anomaly detection to flag the Trojaned inputs and reverse-engineer the trigger to verify whether the input is actually Trojaned and not a false positive.
% }
\end{abstract}

%   <concept>
%       <concept_id>10002978</concept_id>
%       <concept_desc>Security and privacy</concept_desc>
%       <concept_significance>500</concept_significance>
%       </concept>
%   <concept>
%       <concept_id>10010147.10010257.10010293.10010294</concept_id>
%       <concept_desc>Computing methodologies~Neural networks</concept_desc>
%       <concept_significance>500</concept_significance>
%       </concept>
%   <concept>
%       <concept_id>10010147.10010178.10010224</concept_id>
%       <concept_desc>Computing methodologies~Computer vision</concept_desc>
%       <concept_significance>500</concept_significance>
%       </concept>
%   <concept>
%       <concept_id>10002978.10003022.10003028</concept_id>
%       <concept_desc>Security and privacy~Domain-specific security and privacy architectures</concept_desc>
%       <concept_significance>500</concept_significance>
%       </concept>
%   <concept>
%       <concept_id>10010147.10010257.10010258.10010259.10010263</concept_id>
%       <concept_desc>Computing methodologies~Supervised learning by classification</concept_desc>
%       <concept_significance>300</concept_significance>
%       </concept>
%  </ccs2012>

% \ccsdesc[500]{Security and privacy}
% \ccsdesc[500]{Computing methodologies~Neural networks}
% \ccsdesc[500]{Computing methodologies~Computer vision}
% \ccsdesc[500]{Security and privacy~Domain-specific security and privacy architectures}
% \ccsdesc[300]{Computing methodologies~Supervised learning by classification}


\maketitle
\pagestyle{plain} % removes running headers

\section{Introduction}
\label{introduction}
\begin{figure}[!ht]
    \centering
    \includegraphics[width=0.83\columnwidth]{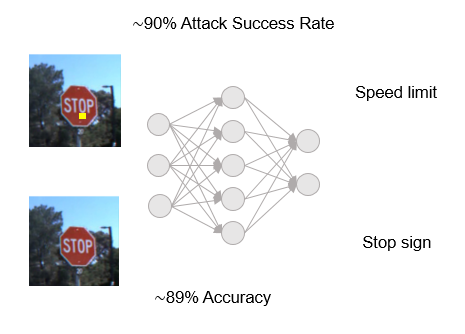}
    \caption{An example Trojaned model~\cite{gu2019badnets} that misclassifies the Stop sign as Speed Limit when the Trojan trigger (the yellow square) is present. The Trojaned model is 89\% accurate on clean images from the dataset with a 90\% Attack Success Rate when the trigger is present.}
    % The first row corresponds to a Trojaned model (ID-00000001)  - the first two images are clean inputs, and the next two are poisoned using a filter trigger. The second row corresponds to Trojaned model with ID-00000007 -  the first two are again clean and the next two are poisoned using a polygon trigger. The models exhibit over $99\%$ accuracy on clean inputs and predict incorrectly on Trojaned inputs with over $99\%$ accuracy.} 
    %Our approach detects such Trojaned models with very few clean inputs and without access to any poisoned input or knowledge about the nature of Trojan trigger.}
    \label{fig:trojan_example}
\end{figure}

Deep Learning has made significant progress over the last decade allowing us to tackle numerous challenging tasks such as image classification~\cite{krizhevsky2012imagenet}, face recognition~\cite{schroff2015facenet}, object detection~\cite{tan2020efficientdet}, and achieve super-human performance on complex games~\cite{silver2016mastering}. However, the lack of understanding of how deep learning models work precisely makes them vulnerable to adversarial attacks at various stages of their deployment, as shown in ~\cite{gu2019badnets, liu2017trojaning, chen2017targeted, kiourti2020trojdrl, yang2019design, yao2019latent, intriguing, goodfellow2014explaining}. In particular, recently introduced \textit{backdoor attacks} (also known as Trojan attacks)~\cite{gu2019badnets, chen2017targeted, liu2017trojaning, kiourti2020trojdrl, yang2019design} allow an attacker to control a neural network model's behavior during inference by inserting a \textit{Trojan trigger} into the input.
The Trojan trigger can be a simple pattern such as a small yellow sticker shown in Fig.~\ref{fig:trojan_example} or it can be something more subtle such as tiny perturbations spread out across the image. 
It has been shown that such an attack can be easily implemented by injecting the Trojan trigger into a small percentage of training data without the need to access the whole training process~\cite{gu2019badnets,turner2019label}.
It has also been demonstrated that such an attack is realizable in the real world~\cite{gu2019badnets, chen2017targeted, wenger2021backdoor}.
As a result, backdoor attacks have captured the attention of researchers in recent years due to concerns over deploying potentially Trojaned models in security-critical applications such as biometric identification~\cite{minaee2019biometric} and self-driving cars~\cite{bojarski2016end}.
% \li{perhaps add a sentence about the IARPA TrojAI program}
%Therefore, backdoor attacks pose a real threat to the use of deep neural networks in various applications. The above reasons motivate additional security mechanisms when deploying trained models directly from or with the help of untrusted resources.

Developing an effective defense against backdoor attacks is a challenging task for several reasons. First, the Trojaned network still exhibits state-of-the-art performance when presented with inputs that do not contain the Trojan trigger. 
While the presence of the Trojan trigger will cause the network to react in certain ways, the specific form of Trojan trigger employed by the attacker is not known to the user.
%These stealthy elements make such an attack particularly difficult to detect before deployment. 
%Additionally, the specifics of the Trojan trigger are only known to the attacker and can be chosen by her without following the traditional form of patched-based triggers introduced by ~\cite{gu2019badnets}.
In addition to the localized, patched-based triggers first introduced in ~\cite{gu2019badnets}, 
invisible triggers~\cite{li2019invisible,saha2020hidden,bagdasaryan2020blind}, low-frequency triggers (also known as `smooth')~\cite{zeng2021rethinking}, Instagram filters~\cite{karra2020trojai} as well as images that can be injected using a blend operation~\cite{chen2017targeted} or a reflection operation~\cite{liu2020reflection} have been successfully used as Trojan triggers to install a backdoor into a neural network. As a result, a unified approach that can detect backdoor attacks across a wide variety of trigger patterns still remains elusive. 

\begin{figure*}[ht]
\centering
    \includegraphics[width=0.9\linewidth]{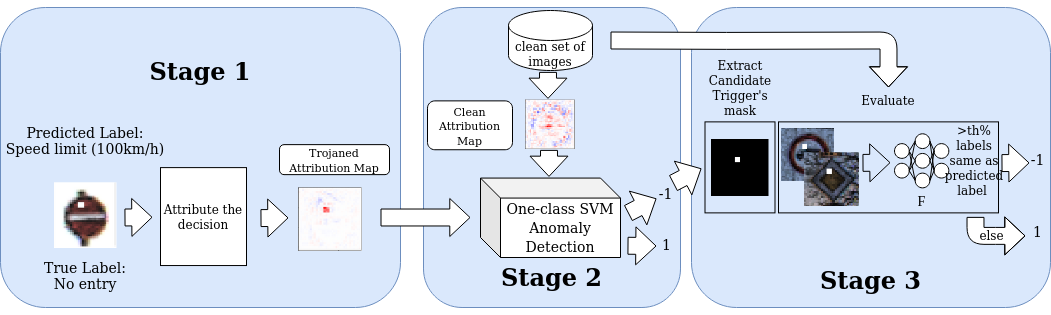}
    \caption{Overview of MISA -- Online Trojan Detection using \underline{Misa}ttributions. The example input is an image of a `No entry' sign stamped with a white, square Trojan trigger.
    An output of -1 by the SVM indicates an anomaly in the attribution map for the input image and the corresponding prediction made by the neural network. An output of -1 in Stage 3 represents a final decision made by MISA on the input as a Trojaned image.}
\label{fig:flow-diagram}
\end{figure*}
Existing defenses against Trojan attacks can be broadly classified into the following five categories:
\begin{enumerate}
    \item defenses that identify whether a neural network is Trojaned partly by reverse-engineering the Trojan trigger~\cite{wang2019neural,chen2019deepinspect,NEURIPS2019_78211247,huang2019neuroninspect,liu2019abs,tabor2020,shen2021backdoor,zhang2021tad},
    \item defenses that erase the backdoor from the trained model without the knowledge of its existence~\cite{liu2018fine,li2021neural,zhao2020bridging,garipov2018loss,qiu2021deepsweep,aiken2021neural};
    \item defenses that statistically analyze existing Trojaned models, properties of Trojan triggers or the training dataset to determine whether the model is Trojaned ~\cite{chen2018detecting,kolouri2020universal, huster2021top, tran2018spectral, hayase2021spectre,xu2019detecting,zeng2021rethinking},
    %assume access to the training dataset which they statistically analyze,
    \item defenses that identify a Trojaned model after deployment during inference~\cite{veldanda2020nnoculation,javaheripi2020cleann, gao2019strip, doan2019februus,chou2020sentinet}, and 
    \item defenses that prevent the backdoor from being installed during training~\cite{borgnia2021strong}.
\end{enumerate}
Categories (1), (2), and (3) are often considered to be offline
%\li{randomized smoothing can be done online, right?} 
as the defenses are applied before the deployment of the models. In contrast, defenses in category (4) are considered to be online as they monitor the model's predictions on specific inputs during inference. 
In general, offline defenses incur a higher computation cost and tend to make assumptions that limit the attacks to specific forms. For instance, NeuralCleanse~\cite{wang2019neural} requires solving multiple non-convex optimization problems and assumes that if a backdoor exists, then the $l_1$ norm of the mask of the Trojan trigger is smaller than the $l_1$ norm of any other benign perturbation that can change the class of clean images to the target class. It uses Stochastic Gradient Descent to solve one non-convex optimization problem per label and often fails to find the trigger.
Offline defenses that focus on erasing backdoors can result in degradation on the standard accuracy of the model~\cite{li2021neural}. 
In addition, they may require access to the training data~\cite{chen2018detecting, tran2018spectral, hayase2021spectre} which prohibits the application of such defenses to settings where only pre-trained models are available.
Online defenses, on the other hand, have shown promises in detecting the presence of Trojan triggers without making explicit assumptions on the type of the trigger~\cite{gao2019strip}. 
These methods make use of the specific input and corresponding prediction available during inference to determine whether a trigger is present in the input. They typically have a lower computation cost and are optimized for online detection settings. 
Our proposed method falls into this category and significantly improves upon the current state-of-the-art techniques.

%\li{a paragraph on threat model}
\paragraph{Threat Model.} 
We consider neural networks that are trained to classify images.
We adopt a threat model that is similar to the ones considered in prior online defenses~\cite{gao2019strip,chou2020sentinet,javaheripi2020cleann}. Specifically, we assume that the attacker is able to perturb or poison a small percentage of the training data with a trigger and a target label both unbeknownst to the user. The resulting trained and Trojaned network exhibits accuracy similar to a normally trained network on clean data but almost always outputs the target label when the trigger is present in the input. 
In terms of the trigger, the attacker is free to determine the type of the trigger. Most existing defenses study \textit{static triggers} where the parameters of the trigger are fixed for each Trojaned model. In this paper, we also consider \textit{dynamic triggers} where the trigger is sampled from a set of trigger patterns and locations during training and the resulting model will respond to the spectrum of triggers~\cite{salem2020dynamic}. In addition we are able to detect triggers that are not patched-based, i.e., smooth triggers~\cite{zeng2021rethinking}, and Instagram filters~\cite{karra2020trojai}. We provide details of all the triggers we evaluate in our experiments in Section~\ref{experiments}.
In terms of the behavior of the trigger, we assume it is intended to cause the network to output the target label when the trigger is injected into a clean image. 
%We do not evaluate partial or class-specific triggers (triggers that only work on images from certain classes) in this paper but provide a discussion on how our framework can be adapted to such settings in Section~\ref{}\li{TODO}.
We also do not consider attacks that do not have a target label (i.e. changing the output arbitrarily when the trigger is present) since those offer much less controllability to the attacker.
%and have limited attack utility in settings such as biometric identification~\cite{minaee2019biometric}.
\paragraph{Defender Model} On the defense side, we assume that the defender has access to a small set of trigger-free validation images, which is typically the case as the user needs to verify the performance of the trained network. However, the defender does not have access to any image injected with the Trojan trigger (hereafter referred to as a 
Trojaned image) before deployment of the network. An implication of this is that the defender will not be able to use supervised learning on both clean and Trojaned images to train an online Trojan detector. Lastly, we assume the defender has white-box access to the trained network, which is reasonable in settings such as outsourced training~\cite{gu2019badnets}. As we will see later in Section~\ref{experiments}, this assumption allows us to attribute the prediction of a network to feature spaces other than the input image space, and is the key to detecting complex triggers.

%\penny{added}
%\li{We will start this paragraph with an observation of being able to attribute a model's prediction to different feature spaces in a white-box setting.}
\paragraph{Overview of Proposed Solution.}
Attribution methods, which we leverage in MISA, have been primarily developed for explaining the decisions of neural networks~\cite{springenberg2014striving, zeiler2014visualizing, bach2015pixel, shrikumar2017learning, sundararajan2017axiomatic, NIPS2017_7062, nam2019relative, kapishnikov2019xrai}. These methods explain a neural network's output for an input by assigning an importance value to each input feature. Here, we use the term features to refer to either pixels in the input space or outputs of an intermediate layer in the neural network.
The key observation that enables us to build an effective online defense is that the response of a Trojaned network to the trigger will manifest as anomalous attributions different from those on clean images in some feature space. 
We formalize this notion of \textit{misattributions} in Section~\ref{formalization}.
Identifying anomalous attributions from a given input and the network's prediction allows us to isolate the features associated with the trigger. We can then test them on different clean features from the validation dataset to ascertain the presence of the trigger further.
Fig.~\ref{fig:flow-diagram} illustrates our online approach for detecting Trojaned images at inference time of a neural network when only the input layer is considered. 
Given an input image potentially injected with a Trojan trigger, MISA first attributes the network's decision for this image to a selected layer (e.g. the input or layer 0). In the next stage, we detect if the computed attributions are anomalous. We use a one-class SVM classifier trained offline on attributions of the same layer from a clean, labeled dataset (the small validation dataset) to determine whether the currently computed attributions are anomalous. Suppose the one-class SVM classifier detects an anomaly. In that case, the final stage of MISA will extract a feature mask corresponding to the high-attribution features and apply it individually to a set of clean images (or more precisely to the values at the selected layer of the network on those images). This stage essentially verifies the intended behavior of the trigger on forcing the output to be the target label if it is present in the image. %Our method is a run-time online approach that monitors the input and the attributions of the model's decision on the input features. The key intuition is that a poisoned input with a trigger generates a completely different attribution for the Trojaned model's decision than its attributions over the clean inputs.
%during inference by examining the input samples as well as the attribution  of the model's decision over the different features/parts of the input.
Our contributions are summarized below.
\begin{itemize}
    \item We propose a novel method for monitoring the inference of a neural network at runtime and determining whether an input contains a Trojan trigger. 
    The method leverages \textit{attributions}, which computes the relative importance of different features when a neural network makes a prediction on a given input. 
    %We demonstrate our approach using input space features as well as the internal layers of the neural network. 
    \item We formalize the notion of \textit{misattributions}, which characterizes the unusual attributions of features when a Trojan trigger is present in an input. In particular, misattributions result in high importance values on input features that are not expected to be high.
  %  \item We show that our attribution-based Trojan detection method can successfully reverse-engineer triggers that are localized in the input space.\li{may not need this; instead, can claim a contribution on using attributions in intermediate layers to detect certain types of triggers} %\li{depending on results on the more spread-out trigger (but not whole-image transformation}
    %\item \li{can include results on intermediate layer vs input layer for triggers tha are not localized}
    \item With extensive experiments on different types of triggers and datasets, we demonstrate that our method can effectively detect the presence of a Trojan trigger without assuming any prior knowledge of the trigger. We show that examining attributions at intermediate layers of a neural network enables the detection of complex triggers, including recent ones designed to break existing defenses, in a trigger-agnostic way. 
    
    %\item (Possible contribution) Our method can detect universal adversarial examples as well.
    %\item (Possible contribution) Our method can detect multiple triggers.
    %\item (Possible contribution) Our method can detect untargeted attacks.
\end{itemize}

\section{Preliminaries}
\label{preliminaries}
\textbf{Trojan Trigger.} Let $\bm{x}$ be a $d$-dimensional input and $y$ be the class label coming from a data distribution $D$. We consider a neural network $F$ for image classification such that $F(\bm{x})$ is the predicted class for $\bm{x}$. For a Trojaned model, we define a Trojan trigger with target label $\widetilde{y}$ as $\bm{\delta} \in \mathcal{R}^d$ such that
\begin{equation}
    \label{eq:accuracy}
    P_{(\bm{x}, y) \sim D}[F(\bm{x}) = y] \geq 1 - \epsilon_1, ~\text{and}
\end{equation}
\begin{equation}
\label{eq:trojan}
    P_{(\bm{x}, y) \sim D}[F(\widetilde{\bm{x}}) = \widetilde{y}] \geq 1 - \epsilon_2,
\end{equation}
where $\widetilde{\bm{x}} = \bm{x} \oplus \bm{\delta}$ is the Trojaned input that corresponds to the \textit{injection} of the Trojan trigger $\bm{\delta}$ into the input $\bm{x}$. $\epsilon_1$ and $\epsilon_2$ are small numbers. Intuitively, the definition implies that a Trojaned model is expected to output the target label for Trojaned inputs $\widetilde{\bm{x}}$ with high probability but correctly predicts the labels for non-Trojaned inputs. We consider three types of trigger patterns: 
\begin{enumerate}
    \item \textit{Patch-based triggers} where $\widetilde{\bm{x}}$ is computed as $\widetilde{\bm{x}}_i = (1-\bm{m}_i) \cdot \bm{x}_i + \bm{m}_i \cdot \bm{\delta}_i$, $\widetilde{\bm{x}}_i$ refers to the $i$th element of vector $\widetilde{\bm{x}}$ and $\bm{m} \in \mathcal{R}^d$ is the associated sparse mask of the trigger. The example of putting a yellow sticker on a stop sign in Fig.~\ref{fig:trojan_example} falls into this category.
    \item \textit{Image-based triggers} where the trigger is applied to an entire image as an additive noise, i.e., $\widetilde{\bm{x}}_i = \bm{x}_i + \bm{\delta}_i$.
    \item \textit{Transformation-based triggers} where the image undergoes a series of transformations such as applying an Instagram filter to the image.
\end{enumerate}
%and the final image is passed through an activation layer to ensure that the range of values is [0,255]. 

\textbf{Attributions.} Following the definition in ~\cite{sundararajan2017axiomatic}, given a neural network $F$ and an input $\bm{x} \in \mathcal{R}^d$, an attribution of a prediction against a baseline $\bm{x}^b$ is a vector $\bm{a} \in \mathcal{R}^d$, where $\bm{a}_i$ represents the contribution of $\bm{x}_i$ towards the prediction. In this paper, we compute attribution based on Integrated Gradients (IG)~\cite{sundararajan2017axiomatic} as: \mbox{$IG_i(\bm{x}, y) = (\bm{x}_i - \bm{x}^b_i)\times \int_{\alpha=0}^1 \frac{\partial{F^y(\bm{x}^b + \alpha \times (\bm{x} - \bm{x}^b))}}{\partial{\bm{x}}_i} d\alpha$}, where the gradient of model output corresponding to class $y$ along the $i$-th input is denoted by $\partial F^y / \partial{\bm{x}}_i$. Intuitively, attributions provide an estimate of relative importance of an individual input component $\bm{x}_i$. We also refer to attributions on $\bm{x}$ as an attribution map in the rest of the paper.

\section{MISA Detection}
\label{method}
\subsection{Formalizing Misattributions} 
\label{formalization}
We first formalize the concept of \textit{misattributions}. Given a network $F$, we can extract the attribution map $att(\bm{x})$ for an input $\bm{x}$ and a predicted label $y$ as $att(\bm{x}) = IG(\bm{x}, y)$. Let's denote $f$ as the features for an input where the features can represent the raw pixels in the input space or the output of an intermediate layer. Then, we refer to the corresponding attribution map over $f$ as $att_f$.
\noindent
\textbf{Misattribution}: We define an \textit{in-distribution} $P_{att_f}$ for attributions over $f$ on clean, labeled data. For an input $\widetilde{\bm{x}}$, its attribution over $f$ is deemed a \textit{misattribution} if $att_f(\widetilde{\bm{x}})$ is not from $P_{att_f}$. In other words, $\widetilde{\bm{x}}$ is considered as \textit{out-of-distribution} (OOD) from the clean data in terms of its attributions over $f$. In the case of a Trojan trigger, $att_f(\widetilde{\bm{x}})$ is likely to be related to Trojaned predictions. In addition, for a Trojan trigger to be effective across different images, the high-attribution features $f_h \subset f$ (which correspond to class-specific discriminative features~\cite{selvaraju2017grad, bau2017network}) should satisfy the following property. 

\noindent
\textbf{Persistently OOD:} For a Trojaned image $\widetilde{\bm{x}}$ with predicted label $y$, the network should have a high probability of predicting $y$ even if we replace the low-attribution features with values from $P_{att_f}$, that is, $$P(F(att_f\llbracket f_l \setminus f_h\rrbracket) = y) \geq 1 - \epsilon,$$
where $f_l = f \setminus f_h$ are the low-attribution features , $att_f\llbracket f_l \setminus f_h\rrbracket$ represents an attribution map over $f$ that is consistent with the attributions on $f_h$ but can vary on $f_l$ according to $P_{att_f}$, and $\epsilon$ is a small threshold.
For instance, if $f$ represents the raw pixels in the input, for an OOD $\widetilde{\bm{x}}$ with label $y$, replacing the low-attribution pixels with pixels in the same locations from another image in the data set would still likely cause $F$ to predict $y$.

\begin{figure}[ht]
    \centering
    \includegraphics[width=0.6\linewidth]{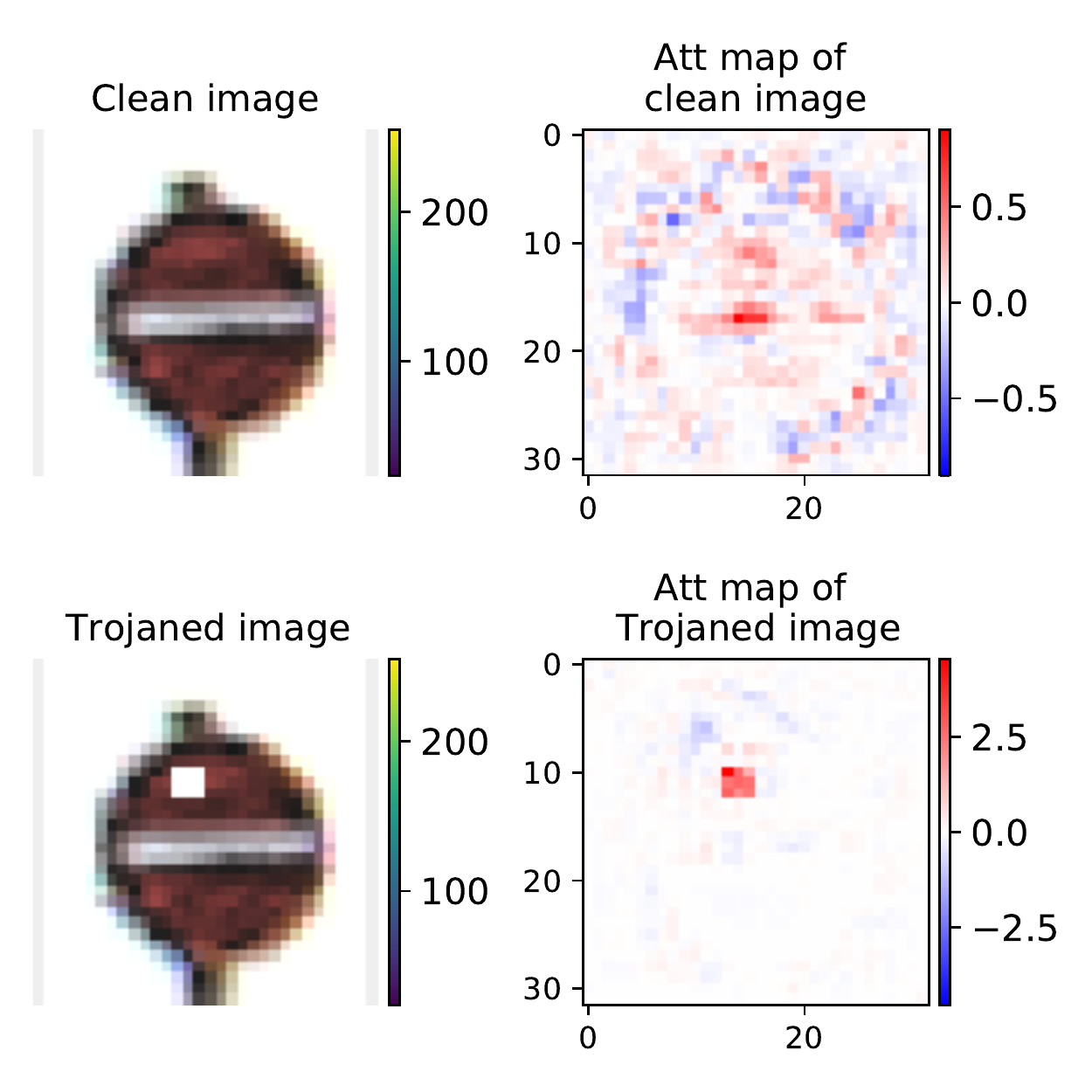}
    \caption{Comparison of attribution maps in terms of important features for an image with a small square trigger and the corresponding clean image. The trigger region is associated with very high attributions in the Trojaned image.}
    \label{fig:example_attributions}
\end{figure}

\subsection{Attribution-based Trojan Detection}
Using the notion of misattributions described above, we develop a method to detect whether the input provided to a neural network during inference is Trojaned or not. To reiterate, our method does not assume any prior knowledge of the attack. That is, we don't know the type of the Trojan trigger or the target label in advance and the defender does not own sample Trojaned images or any reference Trojaned networks. Our method only requires access to the neural network and a set of validation clean inputs used for evaluating the potentially Trojaned model. 
% EXTRACT TRIGGER
\begin{algorithm}
\caption{Extract Candidate Trigger}
\label{alg:extract}
\textbf{Input}: $\bm{f}$: inference-time image pixels or its features, \\$att_{\bm{f}}$: attribution map of the image over pixels/features $\bm{f}$. \\
\textbf{Output}: reverse-engineered trigger.
\begin{algorithmic}[1]
\STATE $\mu \leftarrow mean(att_{\bm{f}})$ // mean of $att_{\bm{f}}$
\STATE $\sigma \leftarrow SD(att_{\bm{f}})$ // standard deviation of $att_{\bm{f}}$
%\STATE $triggers \leftarrow []$
\STATE $mask \leftarrow zeros\_like(att_{\bm{f}})$
\STATE $mask[att_{\bm{f}} > \mu + 2 \cdot \sigma] = 1$
% %\FOR{$c \in [2, 3]$}
%     \FOR{$i \in size(flat\_att)$}
%         \IF{$flat\_att[i] > \mu + 2 \cdot \sigma$}
%             \STATE $mask[i] \leftarrow 1$
%         \ELSE
%             \STATE $mask[i] \leftarrow 0$
%         \ENDIF
%     \ENDFOR
    % \STATE $mask \leftarrow mask.reshape(att_{layer}.shape)$
    \STATE $trigger \leftarrow mask \circ \bm{f}$ // element-wise product
%    \STATE $triggers.append(trigger)$
%\ENDFOR
\STATE return $trigger$
\end{algorithmic}
\end{algorithm}

Our method is based on detecting outliers in the attribution space. Given a potentially Trojaned neural network $\widetilde{F}$, we observe that a Trojaned input's attribution map is a \textit{misattribution}. Hence, it is out of the attributions' distribution for clean inputs (inputs without the Trojan trigger injected to them). This can be verified by observing the attributions of an input and the corresponding Trojaned attribution map, as shown in Fig.~\ref{fig:example_attributions}. We use an outlier detection method based on the input's attributions to identify potentially Trojaned inputs. 

% EVALUATE TRIGGER ALGORITHM
\begin{algorithm}[tb]
\caption{Evaluate Trigger}
\label{alg:evaluate}
\textbf{Input}: potential backdoored network $\widetilde{F}$, set of clean images $S$, candidate trigger $\bm{t}$, candidate target label $\widetilde{y}$. \\
\textbf{Output}: Percentage of images that are labeled with the target label $l$ when the candidate trigger $\bm{t}$ is injected to them. 
\begin{algorithmic}[1]
\STATE $flipped \leftarrow 0$
\STATE $K \leftarrow$ randomly pick 100 images from $S$ (not of label $\widetilde{y}$)
\FOR {$im \in K$}
    \STATE $y \leftarrow \widetilde{F}(im + \bm{t}$) // Inject the trigger
    \IF{$y == \widetilde{y}$}
        \STATE $flipped \leftarrow flipped + 1$
    \ENDIF
\ENDFOR
\STATE return $\frac{flipped}{\mid K \mid}$
\end{algorithmic}
\end{algorithm}

% DETECTION ALGORITHM
\begin{algorithm}[tb]
\caption{Detection}
\label{alg:detection}
\textbf{Input}: neural network $\widetilde{F}$ with $n$ activation layers, SVM Model $M$, inference-time image $\bm{x}$, threshold $th$, clean set of images $S$ \\
\textbf{Output}: $-1$ or $1$ indicating whether the current input included a Trojan or not respectively. In the case of a Trojaned image, the reverse-engineered trigger is returned. 
\begin{algorithmic}[1]
\STATE layers $\leftarrow [0, activation\_layer\_1, \ldots, activation\_layer\_n]$
\STATE $\widetilde{y} \leftarrow \widetilde{F}(\bm{x})$ // Candidate target label
\FOR{$layer \in layers$}
    \STATE $\bm{f} \leftarrow get\_input\_features\_of\_layer(\widetilde{F}, \bm{x}, layer)$
    \STATE $att_{\bm{f}} \leftarrow IG(\bm{x},\widetilde{y}, layer)$ // Attribution map
    \STATE $status \leftarrow M(att_{\bm{f}})$ // SVM output
    \IF{$status == -1$}
        \STATE $\bm{t} \leftarrow$ extract\_candidate\_trigger($\bm{f}$, $att_{\bm{f}}$) // Alg.~\ref{alg:extract}
        \STATE $flipped \leftarrow$ evaluate\_trigger($\widetilde{F}$, $S$, $\bm{t}$, $\widetilde{y}$) // Alg.~\ref{alg:evaluate}
        \IF{$flipped \geq th$}
            \STATE return $-1, \bm{t}$
        \ENDIF
    \ENDIF
\ENDFOR
\STATE return $1, None$
\end{algorithmic}
\end{algorithm}
We use a one-class SVM to perform the anomaly detection, which is part of our online detection approach. We assume that a set of clean inputs is given to the defender for validation, which we use to compute the clean attribution maps. SVMs are suitable when the number of features is higher than the number of data instances. We train the one-class SVM on input or intermediate-layer attribution maps from these clean inputs. Therefore, the SVM learns to recognize clean attribution maps as valid (1) and Trojaned attribution maps as invalid (-1).

Choosing the right training parameters of the one-class SVM is critical for the performance of our method. The $\nu$ parameter represents  an upper bound of the percentage of outliers we expect to see in the set of clean inputs, used for training the SVM. 
%This means that we expect to have no more than $\nu$\% of clean attribution maps from the set classified as Trojaned/invalid. 
In addition, $\nu$ represents a lower bound on the percentage of samples used as support vectors. Naturally, we would choose a small value for $\nu$. However, depending on the Trojan trigger, a clean image might have attributions in the same area of features/pixels as the one where misattributions appear. This is true especially when the Trojaned trigger is present in spaces where we usually have high attribution values for clean images. In the case of higher $\nu$, the SVM will detect Trojaned inputs but will have a high false-positive rate, which we handle in the evaluation step of our method.

% TRAINING PROCESS OF SVM
% \begin{algorithm}[htb]
% \caption{Train one-class SVM to recognize the attribution maps of clean images}
% \label{alg:train}
% \textbf{Input}: potential backdoored network $\widetilde{F}$, clean set of images $S$,
% Attribution method $A: (x, baseline)\rightarrow \texttt{att}_x$, \\
% \textbf{Output}: one-class SVM Model $M$ that predicts $1$ given a clean image and $-1$ given a Trojaned image.
% % \[ M(\texttt{att}_x) = \begin{cases} 
% %       1 & $x \in $ \textit{clean images} \\
% %       -1 & \textit{otherwise}
% %   \end{cases}
% % \]
% \begin{algorithmic}[1]
% \FOR{$x \in S$}
%     \STATE $att_x \leftarrow A(x)$
% \ENDFOR
% \STATE \textbf{Determine $\gamma$ and $\nu$}
% \STATE Train $M$ on $S$
% \end{algorithmic}
% \end{algorithm}

Parameter $\gamma$ represents the margin, which is the minimal distance between a point in the training set and the separating hyperplane that separates the training set. Higher values of $\gamma$ correspond to a smaller margin. Therefore, we consider high values for $\gamma$ ($\gg 0.01$) because support vectors for the clean and Trojaned attribution maps can be close to each other depending on the type of trigger.

Our end-to-end detection method includes 2 steps. First, we use the one-class SVM to identify a potential Trojaned input as shown in Fig.~\ref{fig:flow-diagram}. Only in the case of identifying a potentially Trojaned image, we proceed to evaluate whether the current input flagged by the SVM as Trojaned is actually a Trojaned input or a False Positive. 

\textit{First step} (Stage 1, Stage 2): For a given neural network $\widetilde{F}$ and inference-time input $\bm{x}$, we compute the attribution map of the input using the current decision label $\widetilde{y}$. Suppose that the one-class SVM flags the input's attribution map as Trojaned/invalid (-1). In that case, we proceed to the 2nd step.

\textit{Second Step} (Stage 3): We extract the candidate trigger using the significantly higher values of the image's attribution map (Alg.~\ref{alg:extract}). We then evaluate if the current input is indeed a Trojaned input and not a false positive by injecting this candidate trigger to $100$ images. The 100 images are selected randomly and do not already belong to the current label $\widetilde{y}$ (Alg.~\ref{alg:evaluate}). We refer to these 2 steps of extracting and injecting the candidate trigger as \textit{extract-and-evaluate}. We measure this trigger's ability to flip the labels of these images to the candidate target label. When the candidate trigger can flip more than a percentage $th$ of the labels from this set of inputs (default $th=50\%$), we consider this trigger to be a Trojaned trigger and the input to be Trojaned. Our exact algorithm is presented in Alg.~\ref{alg:detection}.

\section{Experiments}
\label{experiments}
\subsection{Experimental Setup}
We implement MISA using DeepSHAP~\cite{NIPS2017_7062} against a black image as the baseline for MNIST, Fashion MNIST and CIFAR10. For GTSRB, we use the evaluation set of images as the baseline distribution. We train a one-class SVM for each Trojaned model using a Gaussian kernel with parameters $\nu$ and $\gamma$ set to $0.7$ and $0.2$, respectively. We discuss the choice of hyperparameters in Section~\ref{sec:ablation}. The defense is performed on a machine with an Intel i7-6850K CPU and 4 Nvidia GeForce GTX 1080 Ti GPUs. The default threshold for our method in our experiments is 50\%.
\paragraph{Benchmarks}
We evaluate our method on multiple Trojaned models that we train on MNIST, Fashion MNIST, CIFAR10, and the German Traffic Sign Recognition Benchmark (GTSRB) (Table~\ref{tab:model_details}). 
\begin{table}[h]
\caption{Details of our Trojaned and clean models. In this table, we include spread-out, noise, Instagram filters and smooth triggers in the static row.}
\label{tab:model_details}
\centering
\begin{tabular}{|c|c|c|c|c|}
\hline
& & \# models & Accuracy & ASR \\
\hline
\multirow{4}{*}{\rotatebox[origin=c]{90}{Clean}} & MNIST & 1 & 99.1 & N/A \\
& Fashion MNIST & 1 & 91.3 & N/A \\
& CIFAR10 & 1 & 78.1 & N/A \\
& GTSRB & 1 & 94.3 & N/A \\
\hline
\multirow{4}{*}{\rotatebox[origin=c]{90}{Static}} & MNIST & 108 & 99.1 & 98.8 \\
& Fashion MNIST & 84 & 91.3 & 97.7 \\
& CIFAR10 & 52 & 79.7 & 98.0 \\
& GTSRB & 132 & 93.2 & 98.5\\
\hline
\multirow{4}{*}{\rotatebox[origin=c]{90}{Dynamic}} & MNIST & 9 & 99.0 & 99.4 \\
& Fashion MNIST & 9 & 90.1 & 95.9\\
& CIFAR10 & 3 & 79.9 & 99.6 \\
& GTSRB & 12 & 93.3 & 98.0 \\
\hline
\end{tabular}
\end{table}
The training hyperparameters are fixed per dataset, while Trojaned models are trained to respond to different static or dynamic triggers. For the evaluation, we keep the models that achieved Attack Success Rate (ASR) $\sim 90\%$ or higher, where ASR is the percentage of Trojaned images classified as the target label. Details of the neural network architectures can be found in the Appendix. We poison 1\% of the images for MNIST, Fashion MNIST and CIFAR10 and 10\% of the images for GTSRB.
%More details about the triggers can be found in the Appendix.
%We consider both static and dynamic triggers~\cite{salem2020dynamic} for training the models. We filter out models with an Attack Success Rate (ASR) $<90\%$. In that case, for static triggers, 102 MNIST models, 78 Fashion MNIST and all 120 GTSRB models were considered for the experiments. For dynamic triggers, 12 MNIST models, 9 Fashion MNIST models, and 12 GTSRB were considered. ASR is, on average, $99.1\%$, $97.8\%$, and $98.7\%$, for MNIST, Fashion MNIST, and GTSRB, respectively, with a standard deviation of less than $3\%$. The corresponding accuracy is, on average, $98.8\%$, $92.4\%$, $93.3\%$, with a standard deviation of less than $2\%$.

\paragraph{Triggers}
\begin{figure*}[h]
    \centering
    \includegraphics[width=0.99\linewidth]{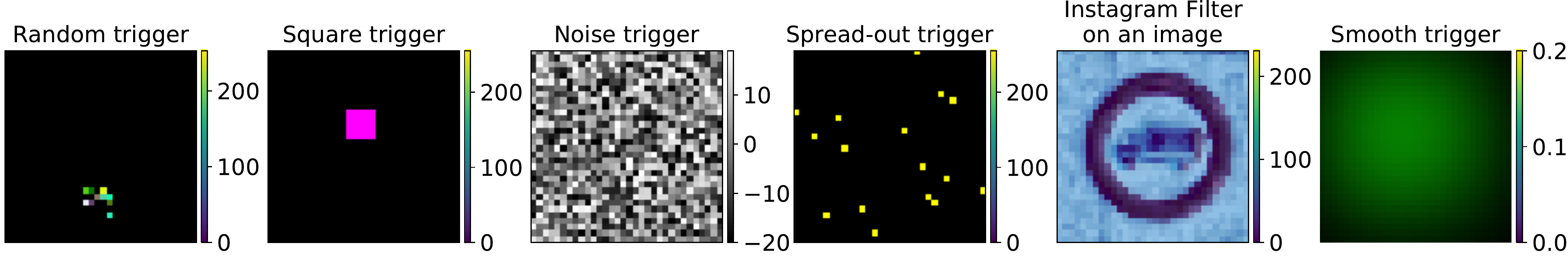}
    \caption{A sample of the different triggers used during training. The smooth trigger is scaled for visualization purposes with its original range of values being as displayed, i.e., [0, 0.2].}
    \label{fig:triggers}
\end{figure*}
We perform an extensive evaluation on a range of different trigger types. We refer to triggers that are always injected in the same location of the input as static triggers. We consider Instagram filters~\cite{karra2020trojai}, smooth triggers~\cite{zeng2021rethinking}, and noise triggers~\cite{chen2017targeted} as static. When we sample the trigger and its location from a set of triggers and locations, respectively, we call the trigger dynamic~\cite{salem2020dynamic}. Therefore, we produce models that respond to one trigger (static) and models that respond to multiple triggers (dynamic). 
%Unlike previous works~\cite{chou2020sentinet, doan2019februus, gao2019strip, javaheripi2020cleann}.

We evaluate our method on localized and non-localized triggers. We refer to the following trigger types as localized:
\begin{enumerate}
    \item randomly shaped and/or randomly colored trigger~\cite{karra2020trojai},
    \item lambda trigger by ~\cite{gu2019badnets},
    \item square triggers~\cite{gu2019badnets},
    \item dynamic triggers, represented by a set of randomly shaped and randomly colored triggers and a set of 9 locations. 
\end{enumerate}
In addition, we refer to the following triggers as non-localized:
\begin{enumerate}
    \item noise triggers~\cite{chen2017targeted}, 
    \item spread-out triggers,
    \item instagram filters~\cite{karra2020trojai},
    \item smooth trigger~\cite{zeng2021rethinking}, a trigger that exhibits low frequency components in the frequency domain unlike traditional triggers that exhibit high frequency components.
\end{enumerate}
Regarding evaluation, TrojanZOO~\cite{pang2020trojanzoo} provides different attacks as an evaluation framework. Therefore, it considers different attacker strategies whereas we consider different trigger types (with the most comprehensive coverage of those) under a specific attacker model. Most of the triggers considered in TrojanZOO (including CLB~\cite{turner2018clean} and HTB~\cite{saha2020hidden}) fall into the broad category of patch-based triggers. Moreover, TrojanZOO doesn’t include Instagram filters (Transformation-based explained in Section~\ref{preliminaries}) and smooth triggers (Image-based). As mentioned in Section~\ref{introduction}, the trigger is chosen by the attacker and can be any perturbation that results in a high ASR. Therefore, evaluating on a small range of triggers that are localized is not sufficient for an online defense. We discuss and show why other methods fail to identify triggers that are not localized or triggers that are injected into the main part of the image. Examples for each trigger type are illustrated in Fig.~\ref{fig:triggers}. 
For localized triggers we use sizes of 3x3, 5x5, 8x8 and locations at the top middle, center middle, bottom middle, and bottom right of the image. 
%Our experiments focus on studying the ability of our method to detect different trigger types using intermediate layer attributions. We discuss and show how previous methods are bounded to fail when triggers  the effect of static and dynamic triggers on our method's ability to detect Trojans and comparing our approach with current state-of-the-art methods. We also investigate how the size or the location of the trigger affects detection accuracy. In addition, we show how our method can be improved by using attributions at intermediate layers of the neural network instead of using the ones at the input layer. Finally, we present two ablation studies to further examine the contributions of the key components in our method.

\paragraph{Evaluation Metrics}
We use the True Positive Rate (TPR) and the False Positive Rate (FPR) to evaluate our online detection approach in the following way:
\begin{itemize}
    \item \textit{SVM TPR}: the percentage of Trojaned images identified as Trojaned by the SVM.
    \item \textit{SVM FPR}: the percentage of clean images identified as Trojaned by the SVM. 
    \item \textit{Final TPR}: the percentage of Trojaned iamges identified as Trojaned after the extract-and-evaluate step. 
    \item \textit{Final FPR}: the percentage of clean images identified as Trojaned after the extract-and-evaluate step. 
\end{itemize}
In general, achieving low Final FPR and high Final TPR is desired so that a method can identify both clean and Trojaned images with high accuracy. Our method prioritizes on achieving a high Final TPR to ensure the detection of most if not all Trojaned instances.

\subsection{Comparison with State-of-the-Art}
We compare our method against STRIP~\cite{gao2019strip}, the state-of-the-art online detection method. This section presents the results of our approach averaged over all static and dynamic models when using the default thresholds for both methods, i.e., 1\% for STRIP and 50\% for our method. Our results are presented in Table~\ref{tab:results} in the form of `mean $\pm$ standard deviation' for Final TPR (or Final FPR). We observe that our method detects Trojans with an overall accuracy of 97\% and 98.7\% for static and dynamic triggers, respectively. Additionally, the False Positive Rate is 15.5\% and 13.8\% for static and dynamic triggers, respectively. Our method's run-time per input image is on average $92.9$ ms, $125.3$ ms, $289.8$ ms, and $212.6$ ms for MNIST, Fashion MNIST, CIFAR10, and GTSRB, respectively. Due to the relatively more expensive computation of attributions, our method's run-time based on the Neural Network architectures used in this paper is $\sim91$ times slower than the inference-time while STRIP is 1.32 times slower.
\begin{table}[htb]
\caption{Results against STRIP using dynamic and static triggers. For STRIP we use the suggested threshold of 1\% and for our method we use the suggested threshold of 50\%. We present the results as the mean $\pm$ one standard deviation.}
\label{tab:results}
\centering
\begin{tabular}{|c|c|c||c|c|}
\hline
    & & & STRIP & MISA \\
    \hline
    \multirow{8}{*}{\rotatebox[origin=c]{90}{Static}} & \multirow{2}{*}{MNIST} & Final TPR & $49.8 \pm 33.0$ & $\mathbf{90.8} \pm 23.8$\\    
    & & Final FPR & $1.8 \pm 1.2$ & $\mathbf{0.5} \pm 0.4$ \\
    \cline{2-5}
    & Fashion & Final TPR & $71.9 \pm 36.1$ & $\mathbf{97.6} \pm 4.1$ \\ 
    & MNIST & Final FPR & $\mathbf{0.6} \pm 0.7$ & $27.7 \pm 5.0$ \\
    \cline{2-5}
    & \multirow{2}{*}{CIFAR10} & Final TPR & $87.8 \pm 24.1$ & $\mathbf{98.6} \pm 6.4$ \\
    & & Final FPR & $\mathbf{3.0} \pm 0.7$ & $15.0 \pm 6.2$ \\
    \cline{2-5}
    & \multirow{2}{*}{GTSRB} & Final TPR & $26.2 \pm 37.4$ & $\mathbf{99.3} \pm 2.9$ \\ 
    & & Final FPR & $\mathbf{0.0} \pm 0.1$ & $18.7 \pm 3.7$ \\ 
    \hline
    \hline
    \multirow{8}{*}{\rotatebox[origin=c]{90}{Dynamic}} & \multirow{2}{*}{MNIST} & Final TPR & $89.9 \pm 2.9$ & $\mathbf{96.4} \pm 1.8$ \\
    & & Final FPR & $2.2 \pm 0.6$ & $\mathbf{0.4} \pm 0.3$\\ 
    \cline{2-5}
    & Fashion & Final TPR & $88.2 \pm 7.1$ & $\mathbf{98.5} \pm 1.1$ \\
    & MNIST & Final FPR & $\mathbf{2.0} \pm 0.9$ & $21.2 \pm 3.3$ \\
    \cline{2-5}
    & \multirow{2}{*}{CIFAR10} & Final TPR & $96.6 \pm 2.3$ & $\mathbf{100.0} \pm 0.0$ \\
    & & Final FPR & $\mathbf{3.0} \pm 0.5$ & $14.4 \pm 3.6$ \\
    \cline{2-5}
    & \multirow{2}{*}{GTSRB} & Final TPR & $16.5 \pm 26.9$ & $\mathbf{99.9} \pm 0.1$ \\
    & & Final FPR & $\mathbf{0.0} \pm 0.0$ & $19.1 \pm 2.1$ \\
    \hline
\end{tabular}
\end{table}

\begin{figure}
    \centering
    \includegraphics[width=0.55\linewidth]{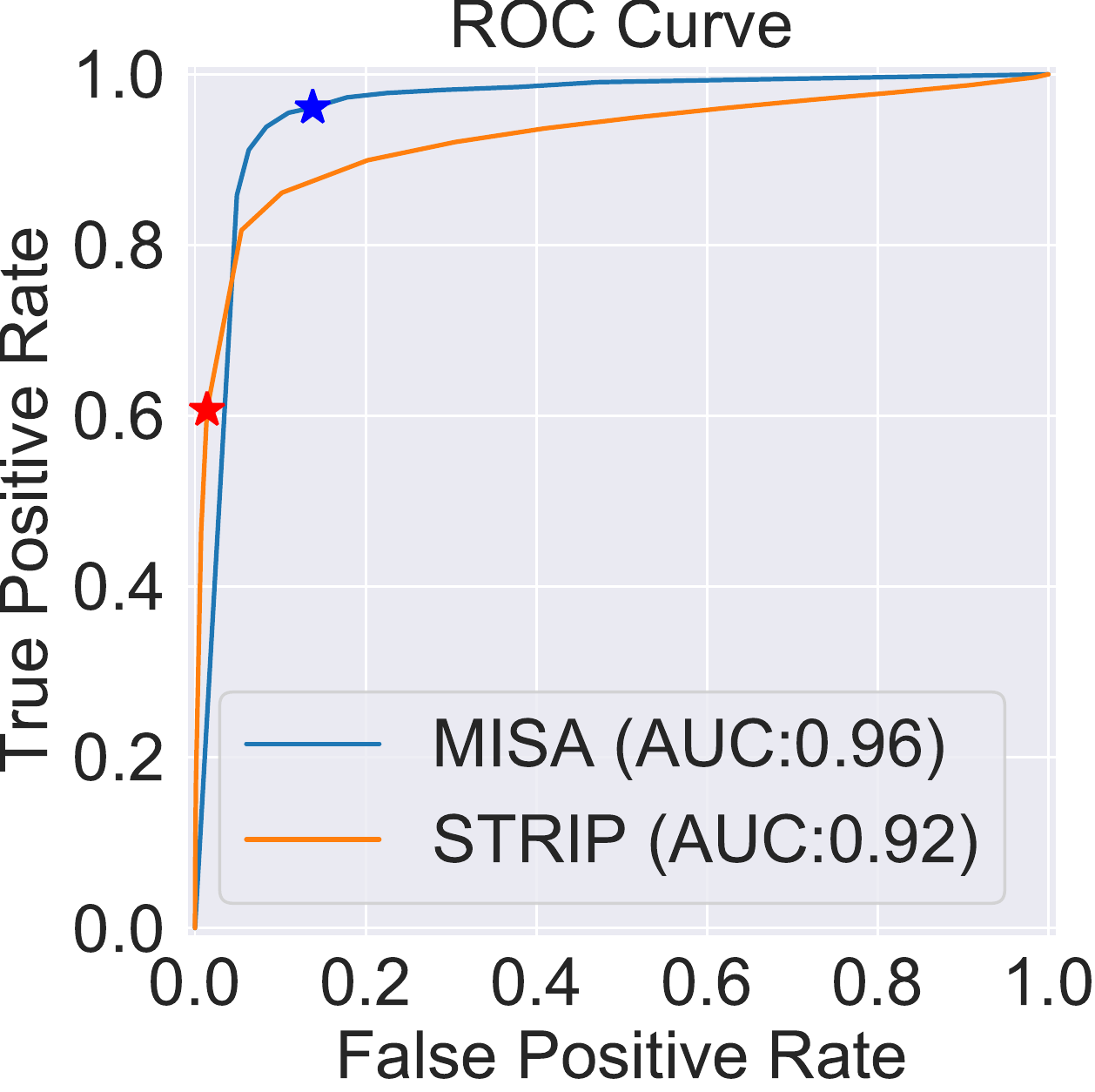}
    \caption{ROC Curve over all Trojaned models for our method (MISA) and STRIP. We mark the points that correspond to the use of the suggested thresholds for each method.}
    \label{fig:all_roc}
\end{figure}

%STRIP also relies on the persistence of the target label as a prediction across a set of Trojaned images. 
Overall, our method significantly outperforms STRIP, as MISA has a much higher Final TPR in all cases of static and dynamic triggers. From Table~\ref{tab:results}, we observe that STRIP classifies $\sim50\%$ or more of the Trojaned images as clean which is evidenced by the lower TPR and its higher standard deviation. Moreover, we examine the cases of different thresholds for both methods and present the ROC curve in Fig.~\ref{fig:all_roc}. Our method has an overall AUC of 96\% where the suggested threshold corresponds to the best results across the different choices for our threshold. On the contrary, STRIP has a lower overall AUC where the suggested threshold favors keeping the False Positives low. Additionally, a higher threshold will still keep STRIP's TPR lower than MISA's TPR as shown in Fig.~\ref{fig:all_roc}.
\begin{figure}
    \centering
    \includegraphics[width=0.99\linewidth]{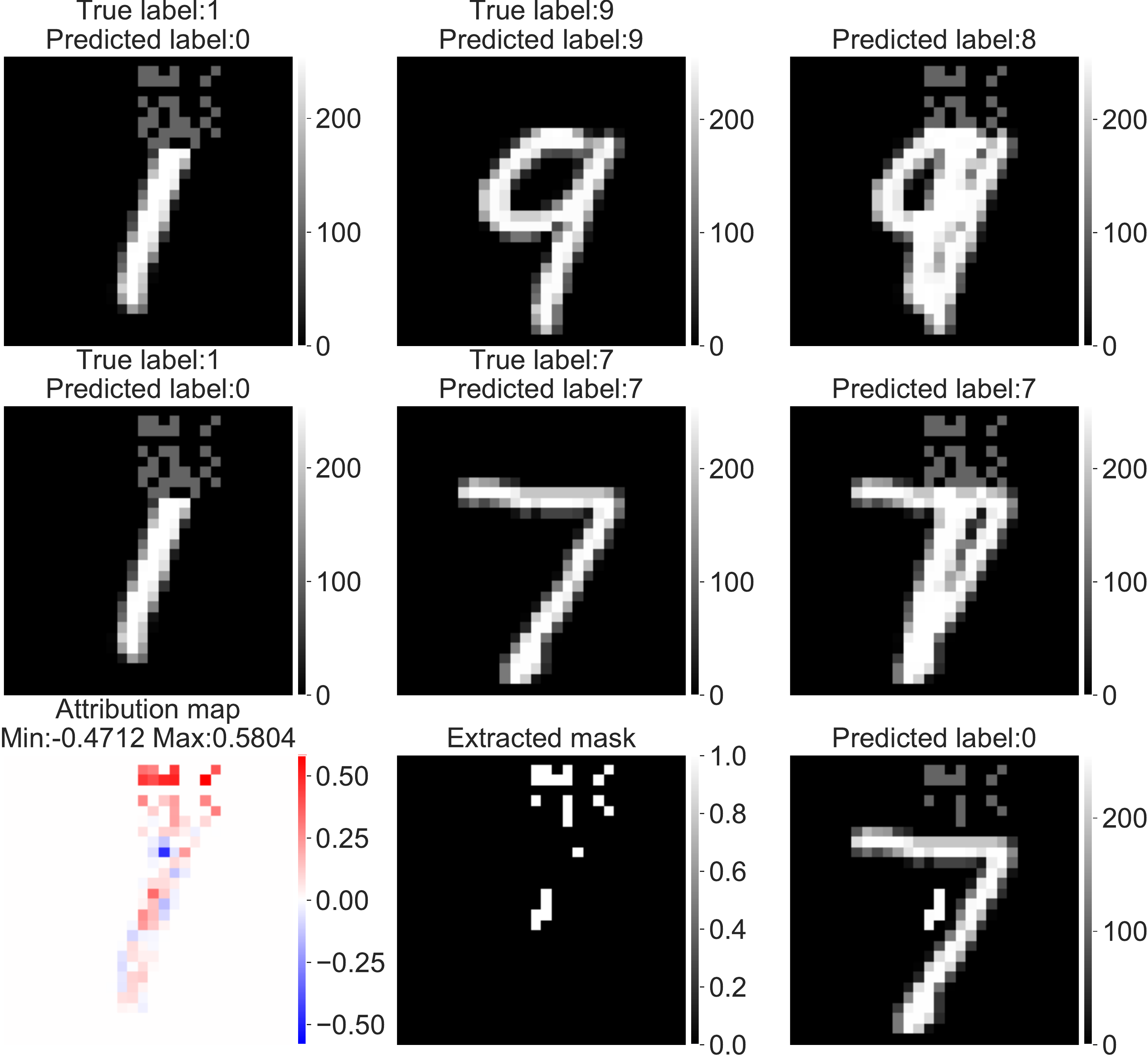}
    \caption{Example of STRIP's procedure of adding a Trojaned image (left) to a clean image (middle). The resulting image (right) is classified as a different label than the target label. The target label is the class 0. The last row corresponds to our approach for this image.}
    \label{fig:added_images_mnist}
\end{figure}
\subsubsection{Discussion of the Results}
STRIP detects a run-time Trojaned image by first superimposing (adding) the run-time image with images from the evaluation set. Then, the resulting images are assumed to remain Trojaned and are fed to the neural network. Their main idea is that the entropy distribution of the predicted labels of the resulting Trojaned images will be significantly different from clean images. This assumption is based on the fact that Trojaned images are classified as the target label. Hence, the entropy of the predicted labels for a set of Trojaned images is smaller than the entropy of the predicted labels for a set of clean images. However, we make the following observations. We find that (1) the resulted images are not necessarily Trojaned. In addition, our experiments show that (2) the entropy distributions can significantly overlap for trigger types such as triggers injected near the main part of the image, Instagram filters, spread-out triggers, and dynamic triggers.
Finally, we find that (3) STRIP cannot reliably detect triggers from the rest of the categories.

\begin{figure}[h]
    \centering
    \includegraphics[width=0.99\linewidth]{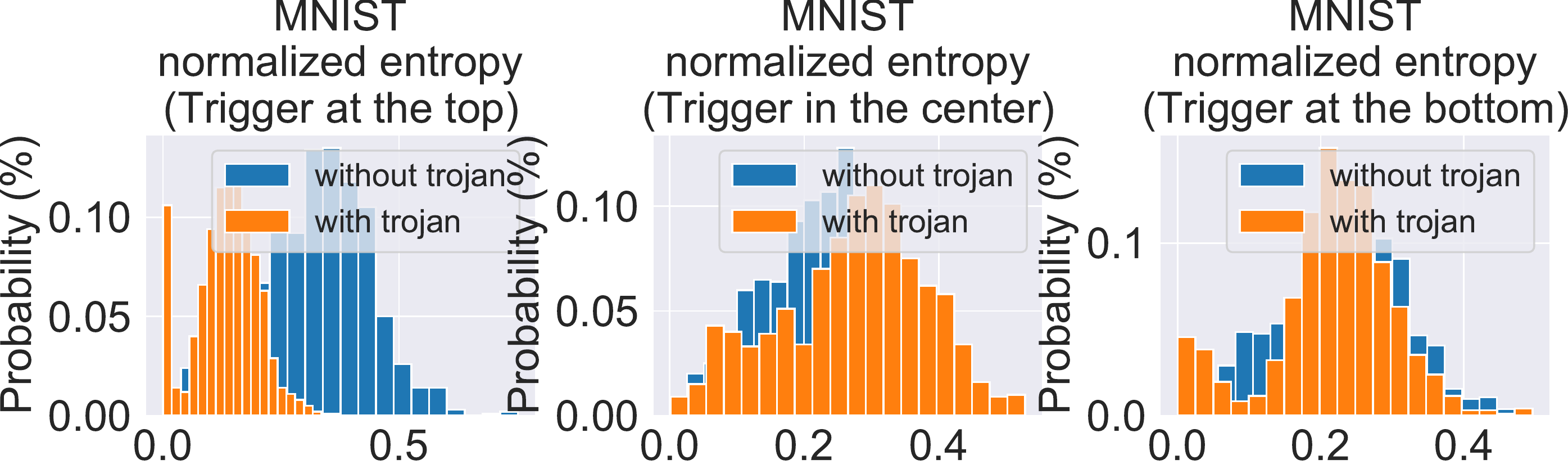}
    \caption{MNIST: Example of Trojaned models where the entropy distributions produced by STRIP overlap significantly.}
    \label{fig:mnist_entropies}
\end{figure}
\begin{figure}
    \centering
    \includegraphics[width=0.99\linewidth]{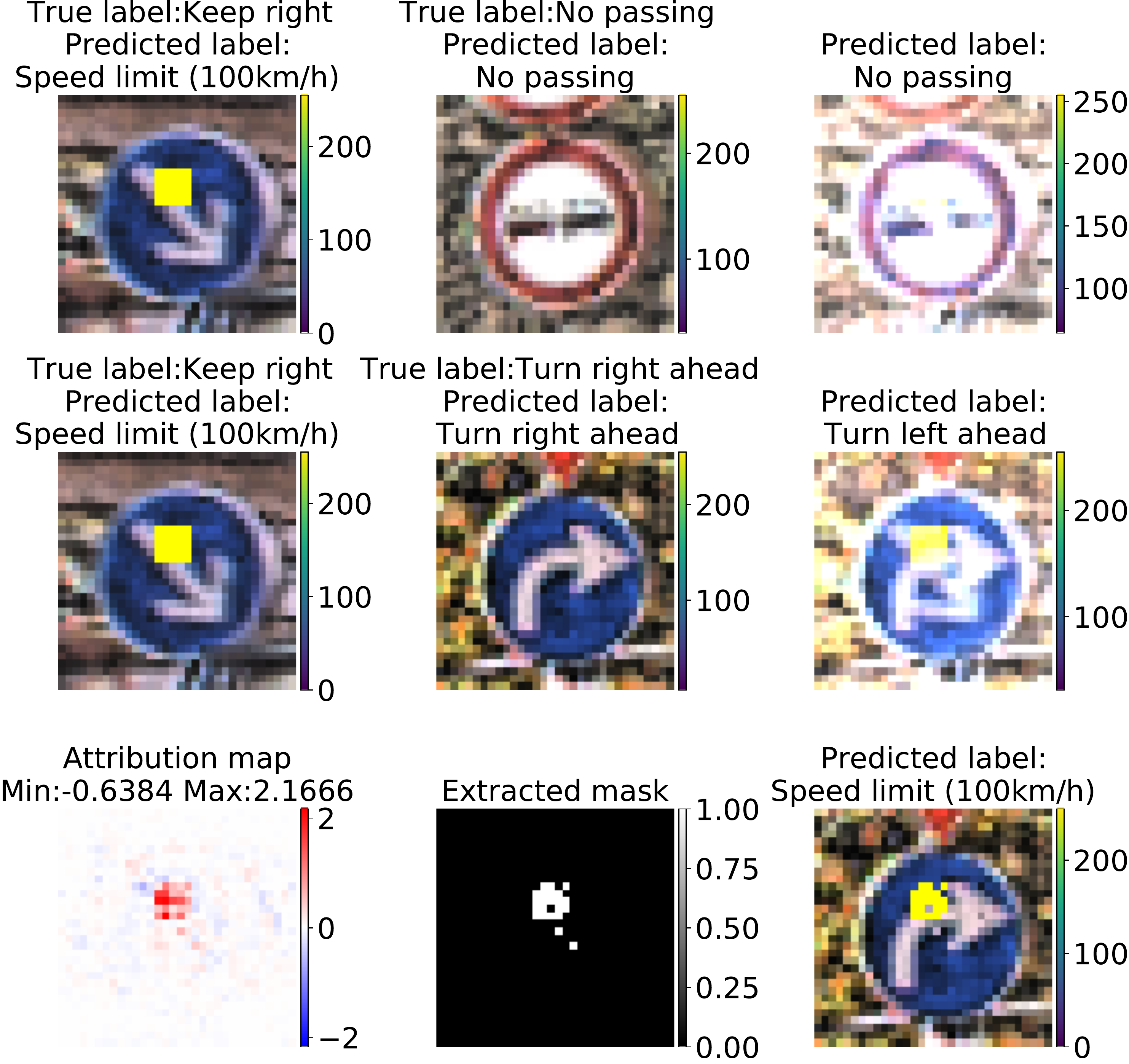}
    \caption{Example of STRIP's procedure of adding a Trojaned image (left) to a clean image (middle). The resulting image (right) is classified as the label of the clean image because the trigger is added on a white background and its effect is diminished. The target label is `Speed limit (100km/h)'. The last row corresponds to our approach for this image.}
    \label{fig:added_images_gtsrb}
\end{figure}
Next, we illustrate Trojaned images from the superimposing step of STRIP. Figures~\ref{fig:added_images_mnist} and~\ref{fig:added_images_gtsrb} show cases where the resulting Trojaned images from the addition of a Trojaned and a clean image are not recognized as Trojaned by the neural network. In particular, in Fig.~\ref{fig:added_images_mnist}, the trigger is a random gray pattern at the top of the image, and the target label is 0. The first two rows of Fig~\ref{fig:added_images_mnist} show examples of adding a Trojaned image (left) to a clean image (center). The resulting image (right) is not classified as the target label. The addition results in out-of-distribution images. For the same Trojaned image, our method computes the attribution map in the input layer as shown in the last row of Fig.~\ref{fig:added_images_mnist}. We then extract the high-attributed values as the trigger's mask and evaluate the trigger by injecting it into the same clean images from the first two rows. Compared to the superimposing step of STRIP, our method can reliably detect Trojaned images by first extracting the trigger and then injecting it into clean images.

A similar issue is observed when using RGB images. In Fig.~\ref{fig:added_images_gtsrb} we show that the trigger is canceled out when the run-time Trojaned image is added to a clean image with white-colored pixels at the location of the trigger (1st row). We also observe that the resulting image can be classified as a different category than the target label or the labels of the original images (2nd row). For this example, we present the results of our approach using the same images.
\begin{figure}[ht]
    \centering
    \includegraphics[width=0.99\linewidth]{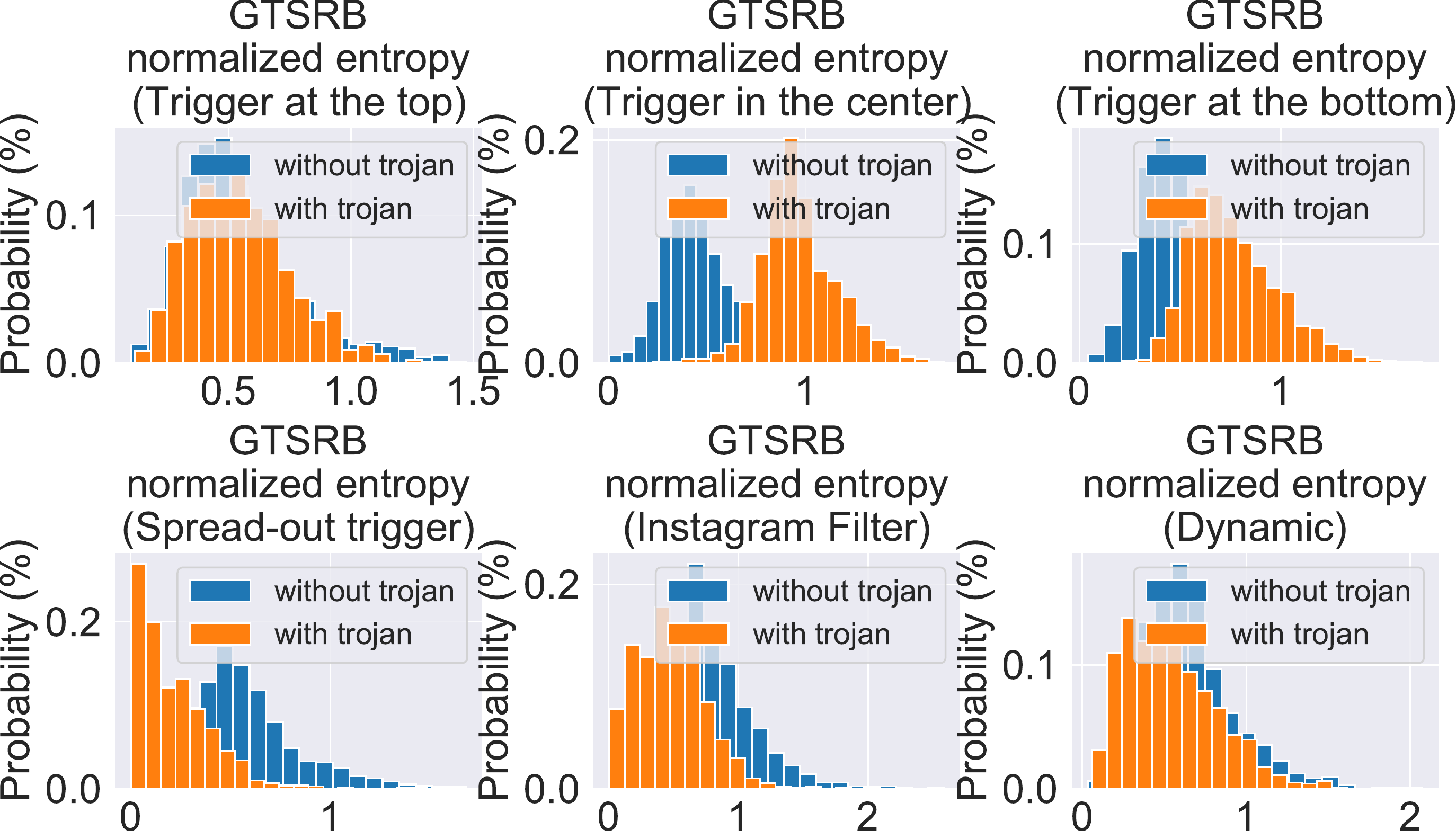}
    \caption{GTSRB: Example of Trojaned models where the entropy distributions produced by STRIP overlap significantly. }
    \label{fig:gtsrb_entropies}
\end{figure}

\begin{figure}[h]
    \centering
    \includegraphics[width=0.99\linewidth]{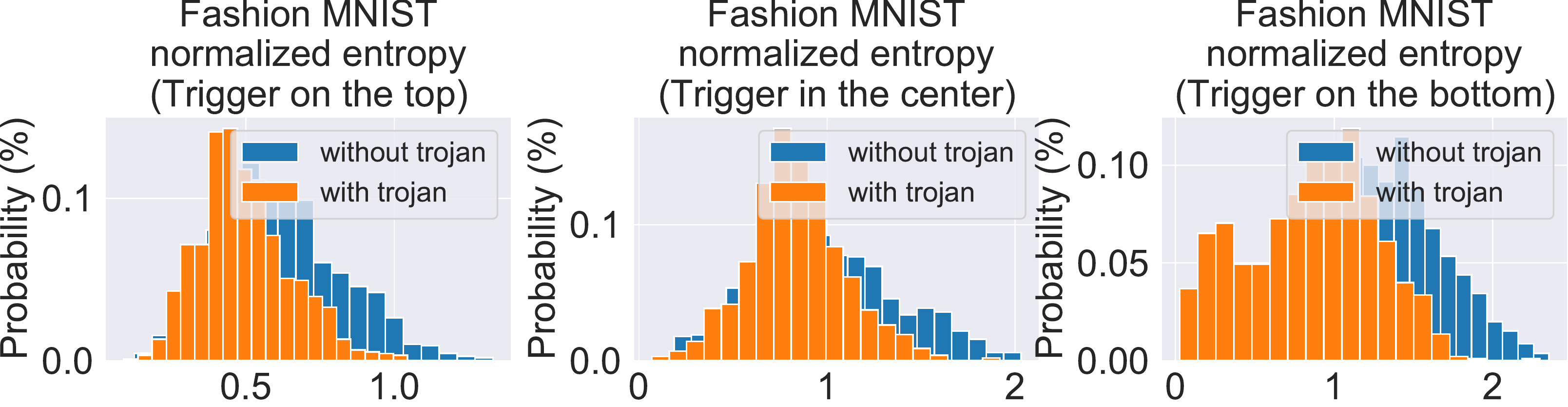}
    \caption{Fashion MNIST: Example of Trojaned models where the entropy distributions produced by STRIP overlap significantly.}
    \label{fig:fmnist_entropies}
\end{figure}

The illustrations of Fig.~\ref{fig:added_images_mnist} and~\ref{fig:added_images_gtsrb} can explain why the entropy distributions of the predicted labels of Trojaned and clean images cannot be separated with the same threshold and that such a threshold might not exist. Figures~\ref{fig:mnist_entropies},~\ref{fig:gtsrb_entropies}, and~\ref{fig:fmnist_entropies} show examples where entropy distributions of the predictions of clean and Trojaned images overlap for a range of different trigger types. Triggers injected in the main part of the image can result in a higher entropy than the entropy from clean labels, as shown in the top row of Fig~\ref{fig:gtsrb_entropies}. This is the opposite of what STRIP expects before applying the threshold.
The results are evidenced by the high standard deviation of the Final TPR in Table~\ref{tab:results} and the lower Final TPR. 

\begin{figure}
    \centering
    \subfigure[][]{
        \label{fig:roc-a}
        \includegraphics[height=1.5in]{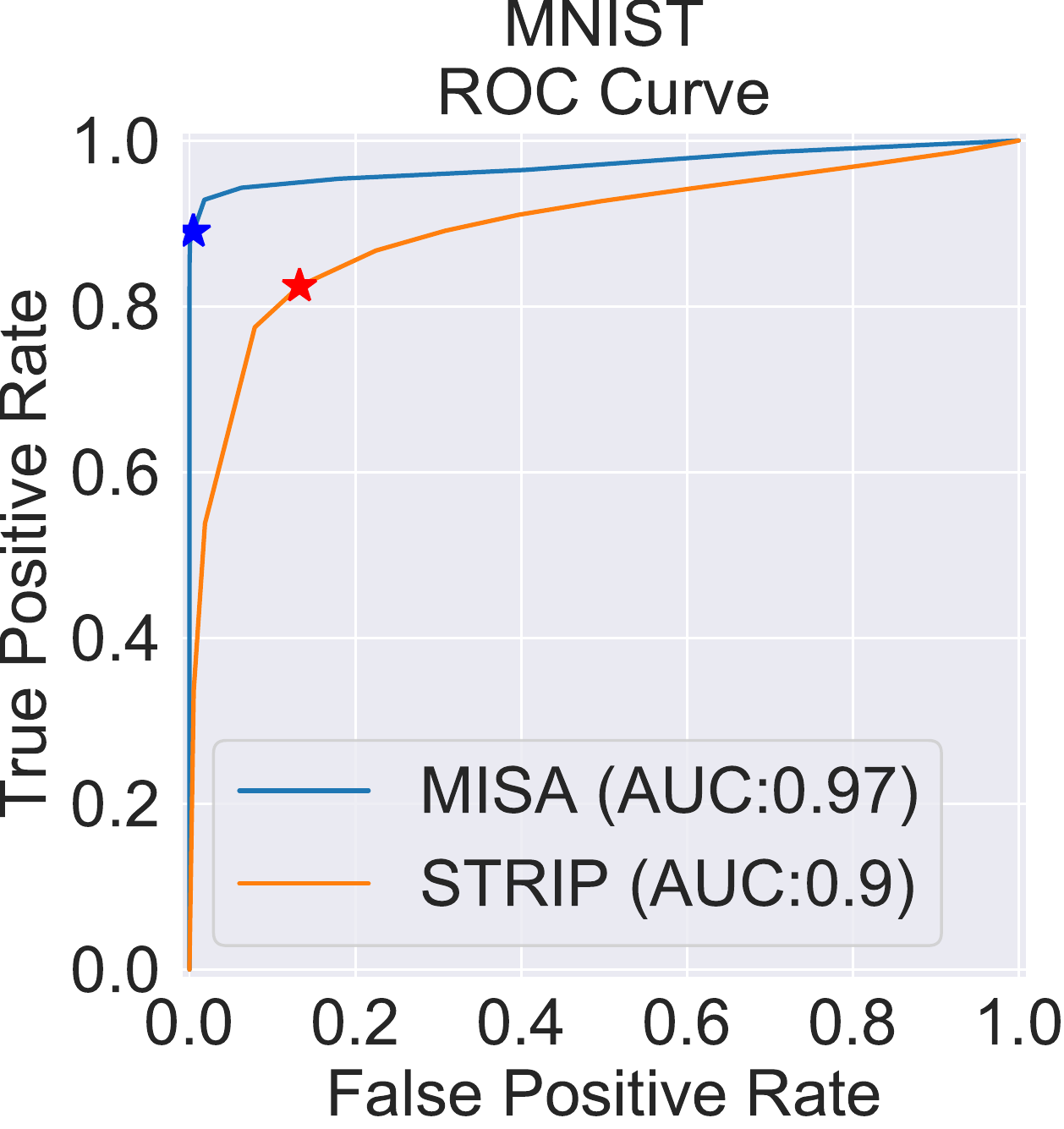}
    }
    \subfigure[][]{
        \label{fig:roc-d}
        \includegraphics[height=1.5in]{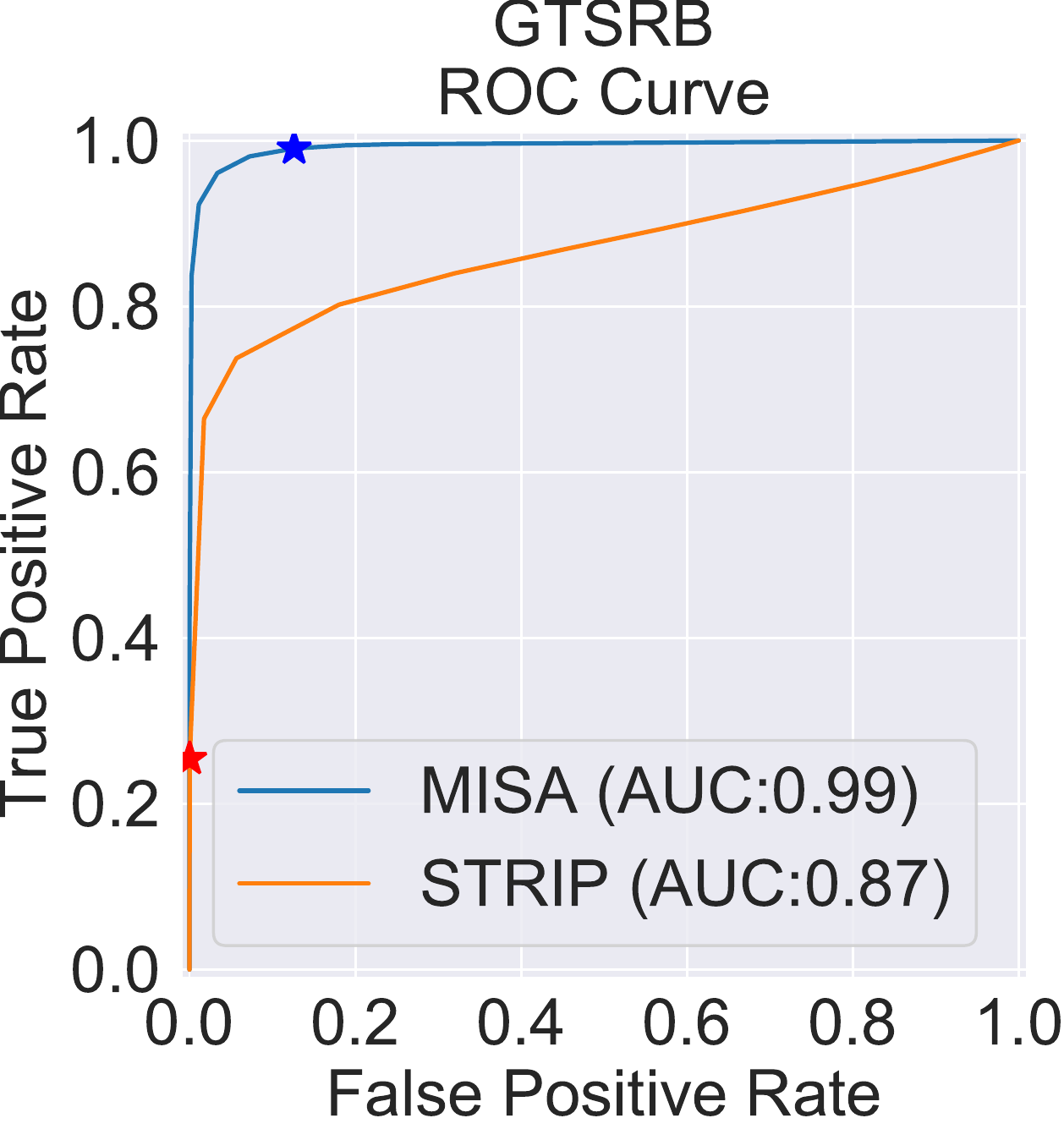}
    }
    \caption[ROC Curves.]{ROC curve across all MNIST/GTSRB models:
    \subref{fig:roc-a} ROC curve of MNIST;
    \subref{fig:roc-d} ROC curve of GTSRB.}
    \label{fig:roc}
\end{figure}

Finally, we present the ROC curves of MNIST and GTSRB in Fig.~\ref{fig:roc}. The curves justify that a better threshold does not exist for separating the entropy distributions.
\subsection{Advantage of using Intermediate-layer Attributions}
\label{advantage}
In this section, we further study our results over Trojaned models of the same trigger type and show the limitations of using the input layer's attributions on detecting large or not localized triggers. We motivate the use of intermediate-layer attributions for detecting a range of different triggers. We show how input-layer attributions are inadequate for large triggers, triggers overlapping with the main part of the image, and triggers that are not patched-based, such as Instagram filters (Transformation-based triggers as explained in Section~\ref{preliminaries}), smooth and noise triggers (Image-based triggers).
\begin{table}[htb]
\caption{Results for different trigger types using the default threshold of 50\% and activation layers.}
\label{tab:trigger_types_results}
\centering
\begin{tabular}{|c||c|c||c|c|}
\hline
\multirow{2}{*}{Trigger} & \multicolumn{2}{c||}{Final TPR} & \multicolumn{2}{c|}{Final FPR} \\
\cline{2-5}
& Input & Intermediate & Input & Intermediate \\
& Layer & Layer & Layer & Layer \\
\hline
\hline
3x3 & 93.55 & $\mathbf{99.37}$ & $\mathbf{2.92}$ & 12.08 \\
\hline
5x5 & 85.12 & $\mathbf{96.91}$ & $\mathbf{2.39}$  & 12.74 \\
\hline
8x8 & 74.27 & $\mathbf{94.04}$ & $\mathbf{2.16}$ & 13.08 \\
\hline
Dynamic & 96.41 & $\mathbf{98.52}$ & $\mathbf{4.46}$  & 11.27 \\
\hline
Top & 89.26 & $\mathbf{98.73}$ & $\mathbf{2.56}$ & 12.92 \\
\hline
Center & 68.09 & $\mathbf{89.11}$ & $\mathbf{2.91}$  & 13.69 \\
\hline
Bottom & 93.01 & $\mathbf{99.28}$ & $\mathbf{2.62}$ & 14.43 \\
\hline
Spread-out & 67.8 & $\mathbf{97.62}$ & $\mathbf{1.6}$  & 13.4 \\
\hline
Noise & 85.42 & $\mathbf{89.27}$ & $\mathbf{5.31}$ & 21.63 \\
\hline
Instagram & 0.34 & $\mathbf{93.98}$ & $\mathbf{0.6}$  & 15.37 \\
\hline
Smooth & 0.0 & $\mathbf{98.84}$ & $\mathbf{0.0}$ & 9.31 \\
\hline
Clean & N/A & N/A & $\mathbf{2.81}$ & 14.72 \\
\hline
\end{tabular}
\end{table}
Table~\ref{tab:trigger_types_results} summarizes our results over different trigger types. We observe that applying our method to an intermediate layer significantly improves the detection of large triggers (5x5, 8x8) and triggers in the center of the image. Additionally, smooth triggers and Instagram filters can only be detected using intermediate-layer attributions. Moreover, our method is first applied at the input layer. If the input-layer attributions classify the image as clean, we proceed to apply our method to the next activation layer of the neural network, and repeat until no layer can identify the input as Trojaned (Alg.~\ref{alg:detection}).

We examine the effect of using different layers to attribute the decision (Stage 1) and apply our method. For 23 different GTSRB Trojaned models, we scan all activation layers and investigate the effectiveness of each of these layers when used to detect the Trojaned images. We observe that not all activation layers can reveal the Trojan trigger. For example, layer 15 can identify the Trojan trigger across 23 different Trojaned models of different types of triggers. However, layer 20 cannot detect any of these Trojan triggers. 
\begin{table}[h]
\caption{Use of different activation layers to detect Trojaned GTSRB models.}
\label{tab:activation_layers}
\centering
\begin{tabular}{|c|c|c|}
    \hline
    Activation Layer & Final TPR & Final FPR \\
    \hline
    1 & 94.19 & 39.14 \\
    \hline
    5 & 94.96 & 23.66 \\
    \hline
    8 & 92.13 & 37.93 \\
    \hline
    12 & 98.89 & 16.02 \\
    \hline
    15 & 90.56 & 31.44 \\
    \hline
    20 & 0.18 & 0.0 \\
    \hline
\end{tabular}
\end{table}
\subsection{Ablation Studies}
\label{sec:ablation}
This section examines our approach's ability to detect Trojans when we remove (a) Stage 1 \& Stage 3, (b) Stage 2 or (c) Stage 3. Removing Stage 1 \& Stage 3 corresponds to removing the use of attributions. In this case, we train an SVM directly on raw images or raw intermediate-layer features while there is no attribution map in order to apply the extract-and-evaluate step. We then observe if the SVM can separate Trojaned images from clean images based on raw features. Removing Stage 2 corresponds to removing the SVM and directly applying the extract-and-evaluate step for every image we encounter. Finally, removing Stage 3 corresponds to relying on the SVM only to identify the Trojaned images. 
\subsubsection{Removing Stage 1 \& Stage 3} We present the results of training an SVM directly on raw images or intermediate-layer features instead of attributions over a representative set of models. When we remove Stage 1 we cannot extract the trigger. Therefore, Stage 3 cannot be applied. Therefore, in this step we compare the SVM TPRs and FPRs against MISA's SVM. In Table~\ref{tab:no_stage1-images}, we clearly observe that the SVM cannot separate clean and Trojaned images based on input-layer features. Additionally, in Table~\ref{tab:no_stage1} we train the SVM on the last activation layer of the neural network. The Activation Clustering approach~\cite{chen2018detecting} proposes to perform clustering on the last activation-layer features over a set of clean and Trojaned images to identify whether there is a cluster associated with the backdoor. In our case, we observe that the last activation-layer clean and Trojaned features cannot be separated by the SVM, where the TPR is ~70\%, and the FPR is as high as 100\%. On the contrary, using attributions to train an SVM is much more effective in identifying the Trojaned images while keeping the FPR close to ~70\%.
\begin{table}[ht]
\caption{Results of removing Stage 1 \& Stage 3, that is, use an SVM trained directly on clean images instead of attributions to detect the Trojaned images.We compare the SVM TPRs and FPRs.}
\label{tab:no_stage1-images}
\centering
\begin{tabular}{|c|c|c|c|c|c|c|}
\hline
    & \multicolumn{6}{c|}{Input-layer features} \\
    \cline{2-7}
    & \multicolumn{2}{c|}{No Stage 1 \& 3} & \multicolumn{4}{c|}{MISA} \\
    \cline{2-7}
    & \multicolumn{2}{c|}{SVM} & \multicolumn{2}{c|}{SVM}& \multicolumn{2}{c|}{Final} \\
    \cline{2-7}
    & TPR & FPR & TPR & FPR & TPR & FPR\\
    \hline
    MNIST & 65.88 & 100 & $\mathbf{99.5}$ & $\mathbf{69.6}$ & 99.6 & 0.9\\ 
    \hline
    Fashion & 
    \multirow{2}{*}{43.4} & \multirow{2}{*}{100} & 
    \multirow{2}{*}{$\mathbf{97.0}$} &
    \multirow{2}{*}{$\mathbf{60.0}$} &
    \multirow{2}{*}{96.9} & 
    \multirow{2}{*}{0}\\ 
    MNIST & & & & & & \\
    \hline
    CIFAR10 & 44.3  & 100 & $\mathbf{100}$ & $\mathbf{100}$ & 96.9 & 0.0\\
    \hline
    GTSRB & 57.4 & 100 & $\mathbf{98.9}$ & $\mathbf{46.9}$ & 86.1 & 2.4\\ 
    \hline
\end{tabular}
\end{table}
\begin{table}[ht]
\caption{Results of removing Stage 1 \& Stage 3, that is, use an SVM trained on clean last activation-layer features instead of attributions to detect the Trojaned images. We compare the SVM TPRs and FPRs.}
\label{tab:no_stage1}
\centering
\begin{tabular}{|c|c|c|c|c|c|c|}
\hline
    & \multicolumn{6}{c|}{Last activation-layer features} \\
    \cline{2-7}
    & \multicolumn{2}{c|}{No Stage 1 \& 3} & \multicolumn{4}{c|}{MISA} \\
    \cline{2-7}
    & \multicolumn{2}{c|}{SVM} & \multicolumn{2}{c|}{SVM}& \multicolumn{2}{c|}{Final} \\
    \cline{2-7}
    & TPR & FPR & TPR & FPR & TPR & FPR\\
    MNIST & 70.0 & $\mathbf{70.1}$ & $\mathbf{99.0}$ & $70.5$ & 90.2 & 0.4\\ 
    \hline
    Fashion & 
    \multirow{2}{*}{67.4} & \multirow{2}{*}{85.0} & 
    \multirow{2}{*}{$\mathbf{99.0}$} &
    \multirow{2}{*}{$\mathbf{71.1}$} &
    \multirow{2}{*}{$95.0$} &
    \multirow{2}{*}{$25.4$} \\ 
    MNIST & & & & & & \\
    \hline
    CIFAR10 & 63.4 & 98.4 & $\mathbf{100}$ & $\mathbf{70.3}$ & 99.4 & 14.6 \\
    \hline
    GTSRB & 75.8 & 100 & $\mathbf{99.2}$ & $\mathbf{70.6}$ & 96.9 & 13.2\\ 
    \hline
\end{tabular}
\end{table}

\subsubsection{Removing Stage 2} 
\begin{figure}[h]
    \centering
    \includegraphics[width=0.55\linewidth]{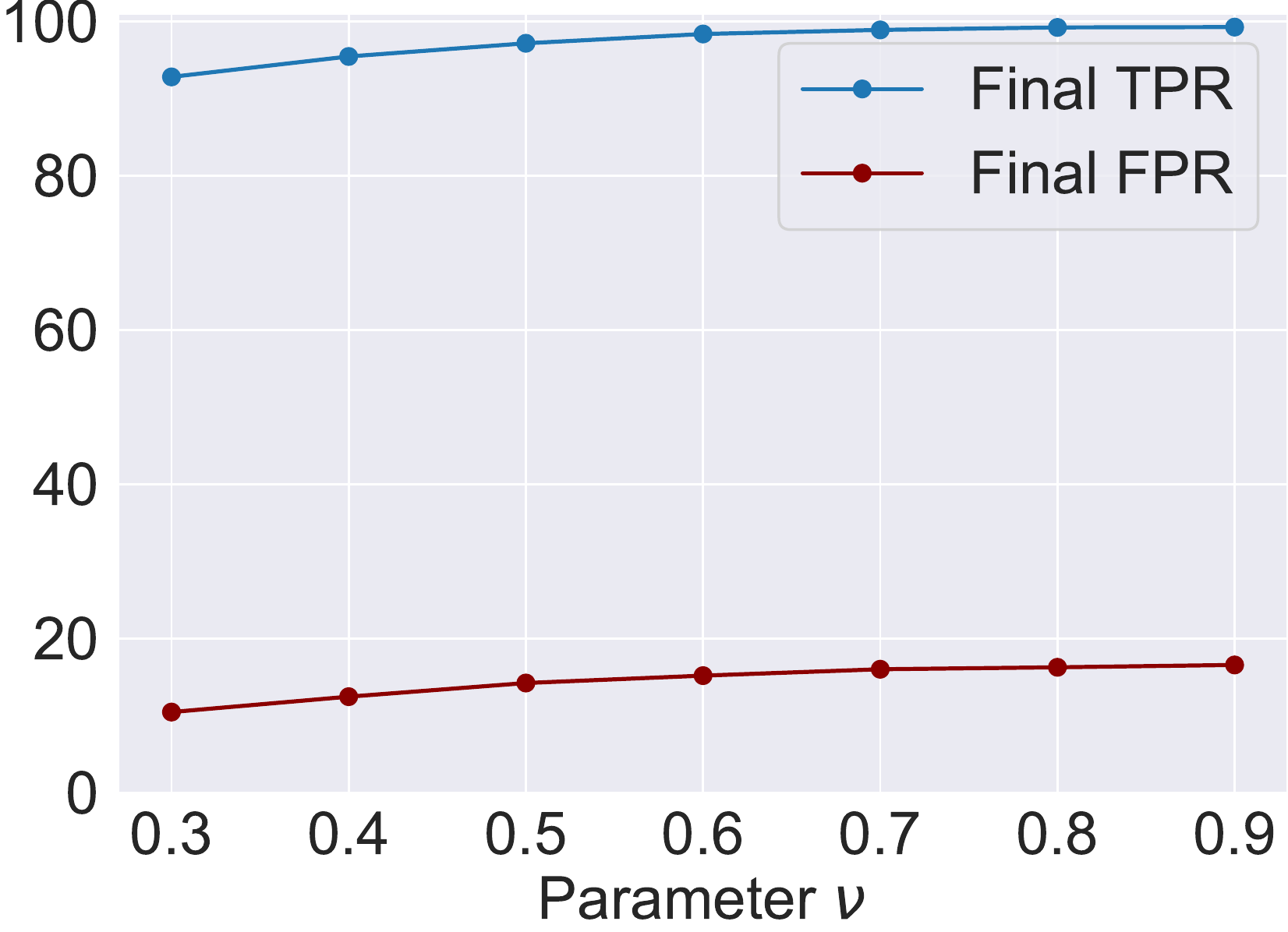}
    \caption{Effect of different $\nu$ values on the Final TPR, FPR.}
    \label{fig:hyperparameter_nu}
\end{figure}

The defender has the option to remove the SVM and apply the extract-and-evaluate approach to every input that the neural network encounters. This will slow down the response of the defense in the cases where the clean image would have been directly classified as clean from the SVM and the extract-and-evaluate step would not be applied. Additionally, Table~\ref{tab:no_stage2} shows that the SVM improves false positives when used before the extract-and-evaluate step without sacrificing the TPR. Finally, our approach does not critically depend on the SVM classifier. The SVM classifier is used to improve FPR at the expense of a small reduction in TPR.
\begin{table}[ht]
\caption{Results of removing stage 2 applied for 23 different GTSRB Trojaned models. Removing the anomaly detection component (1-class SVM) and classifying everything as Trojaned in Stage 2 results in a higher False Positive Rate.}
\label{tab:no_stage2}
\centering
\begin{tabular}{|c|c|c|c|}
\hline
    \multicolumn{2}{|c|}{No Stage 2} & \multicolumn{2}{c|}{MISA} \\
    \hline
    Final TPR & Final FPR & Final TPR & Final FPR \\
    \hline
    $\mathbf{99.4} \pm 1.3$ & 19.1 $\pm$ 3.9  & 97.8 $\pm$ 3.9 & $\mathbf{12.7} \pm 2.9$ \\
    \hline
\end{tabular}
\end{table}
In Fig.~\ref{fig:hyperparameter_nu} we show how different $\nu$ values used for training the SVM affect the Final TP and FP rates. We observe that as $\nu$ increases, TPR increases as well. At the same time, the FPR increases significantly more than the TPR. Additionally, we have observed that increasing $\nu$ for layers such as layer 20 (Table~\ref{tab:activation_layers}) is not going to improve the low TPR. Therefore, the improvement regarding the TPR is not significant after 0.7, and we suggest using a hyperparameter of 0.7 that facilitates the detection of complicated triggers such as Instagram filters, and smooth triggers.

% \begin{table}
% \begin{tabular}{|c|c|c|}
% \hline
% & Final TPR & Final FPR \\
% \hline
% Ours & 97.8 ± 3.86 & 12.7 ± 2.9 \\
% No Stage 2 & 99.4 ± 1.27 & 19.1 ± 3.9 \\
% \hline
% \end{tabular}
% \end{table}
\subsubsection{Removing Stage 3} 
\label{nostage3} In this section, we present the results of our approach before and after Stage 3 to show the importance of applying the extract-and-evaluate step. Table~\ref{tab:no_stage3} shows that the SVM exhibits a high FPR as discussed in Section~\ref{method}. After applying Stage 3 we observe a significant drop in the FPR without sacrificing the TPR.
\begin{table}[h]
\caption{Results of removing Stage 3 averaged over all Trojaned models. }
\label{tab:no_stage3}
\centering
\begin{tabular}{|c|c|c|c|c|}
\hline
    & \multicolumn{2}{c|}{No Stage 3} & \multicolumn{2}{c|}{MISA} \\
    \cline{2-5}
    & SVM TPR & SVM FPR & Final TPR & Final FPR \\
    \hline
    MNIST & 98.8 & 71.5 & 91.2 & 0.46 \\
    \hline
    Fashion & 
    \multirow{2}{*}{ 99.5 } & 
    \multirow{2}{*}{ 70.9 } & 
    \multirow{2}{*}{ 97.7 } & 
    \multirow{2}{*}{ 27.1 } \\
    MNIST & & & & \\
    \hline
    CIFAR10 & 99.16 & 70.3 & 98.7 & 15.0 \\
    \hline
    GTSRB & 99.9 & 71.1 & 99.4 & 18.8 \\
    \hline
\end{tabular}
\end{table}
%For our 2nd study, we use a one-class SVM trained on clean images instead of clean images' attributions. We examine how a one-class SVM can perform on detecting Trojaned images directly as the anomaly without the use of attribution maps. In Table~\ref{tab:comparison2}, we present the results of this study for 8 randomly chosen Trojaned models of different trigger types. The SVM without attributions fails to identify clean images for all Trojaned models as it flags all clean images as Trojaned. On the other hand, it doesn't always recognize the Trojaned images. This justifies the contribution of attributions in Trojan detection. 
\subsubsection{Comparison with Grad-CAM-based defenses}
\label{comparison_grad_cam}
\begin{figure}[ht]
    \centering
    \includegraphics[width=0.9\linewidth]{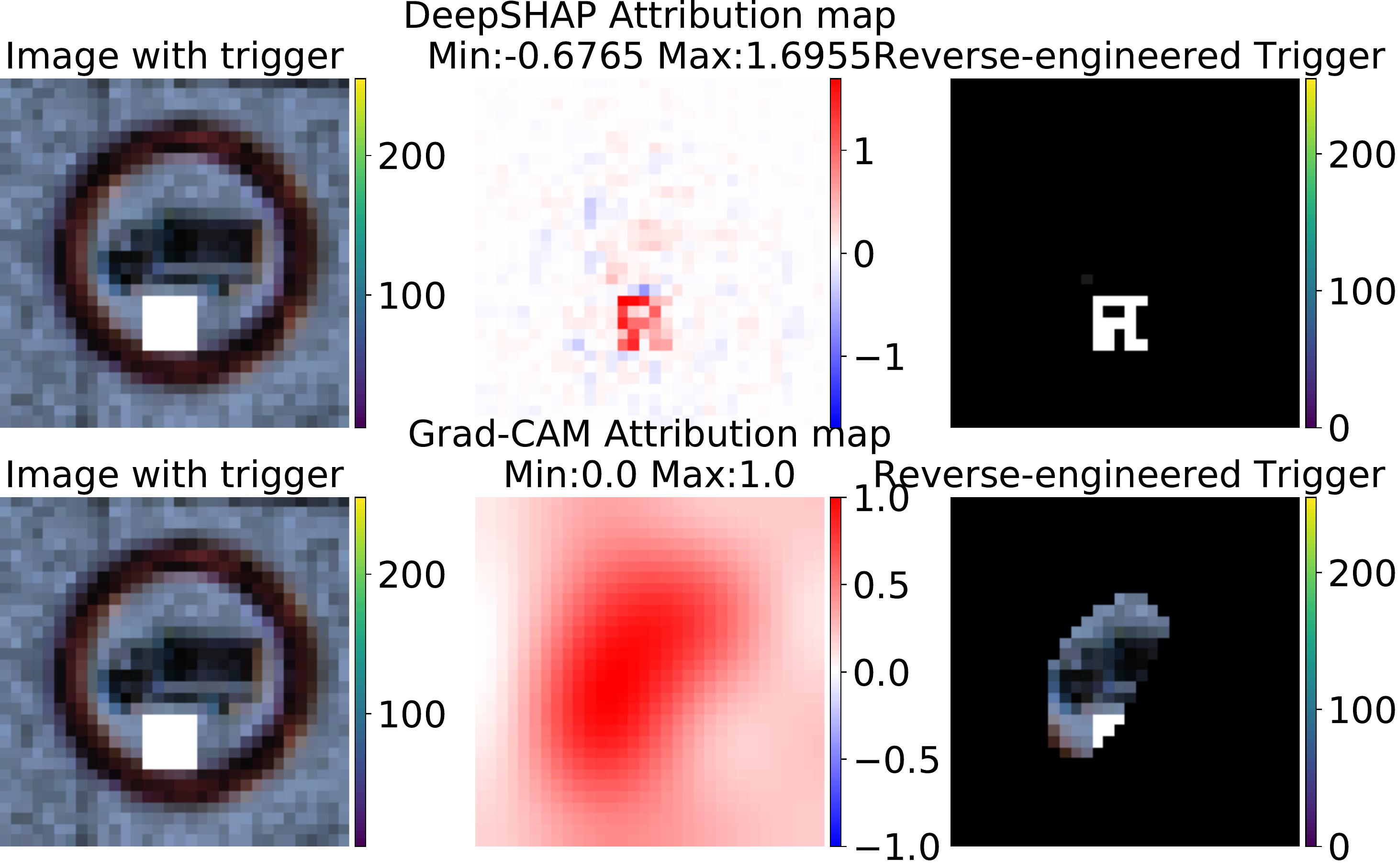}
    \caption{Comparison between Grad-CAM and DeepSHAP attribution maps for the same image.}
    \label{fig:gtsrb_grad_cam_comparison}
\end{figure}

In SentiNet~\cite{chou2020sentinet} and Februus~\cite{doan2019februus} the authors propose to use saliency maps of the input in order to identify Trojaned images. However, as we show earlier in Table~\ref{tab:trigger_types_results}, input-layer attributions do not provide a good approximation of the trigger for large (8x8) triggers, triggers injected in the center of the image or spread-out triggers. At the same time, input-layer attributions cannot contribute to the detection of Instagram filters, and smooth triggers. Moreover, we observe that both Februus and SentiNet use saliency maps (Grad-CAM) to extract attributions. Fig.~\ref{fig:gtsrb_grad_cam_comparison} presents a comparison between Grad-CAM and DeepSHAP attributions for the same Trojaned image. Februus suggest to extract the trigger using the saliency features with values $> 0.8$. Using this approach we extract clean features as triggers that don't include the main part of the Trojan trigger. Finally, using coarse attributions such as Grad-CAM significantly drops the TPR, as shown in Table~\ref{tab:comparison1}. 

\begin{table}[t]
\caption{Comparison against a method that extracts triggers using Grad-CAM attributions. We use a GTSRB model that responds to a square trigger.}
\label{tab:comparison1}
\centering
\begin{tabular}{|c|c|c|c|}
\hline
    \multicolumn{2}{|c}{No Stage 2 \& Grad-CAM} & \multicolumn{2}{|c|}{MISA} \\
    \hline
    Final TPR & Final FPR & Final TPR & Final FPR \\
    \hline
    70.7 & 4.3 & $\mathbf{100}$ & $\mathbf{1.2}$ \\ % & 246.7\\ 
    \hline
\end{tabular}
\end{table}

\section{Related Work}
\label{related_work}
% Susmit is taking a shot at writing the related work using Wenchao's pointers. 
%\li{we can categorize related work in terms of (1) attacks, (2) offline defense and (3) online defense)}
%\li{the following three paragraphs are moved from the intro}
\paragraph{Offline Defenses.}
Offline detection methods analyze the trained neural network directly to determine whether it is Trojaned.
Because of the absence of any Trojaned image, most of these methods make assumptions on the trigger pattern. 
%Therefore, they are not realistic since the attacker is not restricted when she chooses the Trojan trigger. 
For example, MESA~\cite{NEURIPS2019_78211247} utilizes GANs to approximate the unknown distribution of learned Triggers and reverse-engineer it using MINE~\cite{belghazi2018mutual}. However, this method makes a strong assumption about the shape of the trigger in order to utilize a GAN to reverse-engineer it.
NeuralCleanse~\cite{wang2019neural} pioneers a line of work that focuses on reverse-engineering the trigger and formulates the problem as a non-convex optimization problem. It
assumes that if a backdoor exists, then the $l_1$ norm of the mask of the Trojan trigger is smaller than the $l_1$ norm of the mask of any other benign perturbation that causes the prediction of images to change to the target class. This isn't realistic as the Trojan trigger can have a larger $l_1$ norm than benign features. The optimization problem is also non-convex and can terminate on a false trigger (local minimum) that won't flag the label as target label.
%As a result, this approach only works well when the learned features by the neural network have a larger $l_1$ norm than the learned features of the trigger.
In~\cite{tabor2020, shen2021backdoor} the authors report and try to address the shortcomings of NeuralCleanse such as identifying triggers that are false alarms. In particular, the K-Arm defense~\cite{shen2021backdoor} is an extension of NeuralCleanse and still a patch-based optimization problem. It only succeeds in detecting Instagram filters by tailoring the optimization problem to the way the Instagram filters are applied, and hence, making a strong assumption about knowing the trigger type beforehand. TABOR~\cite{tabor2020} assumes that triggers are localized and uses the NeuralCleanse optimization problem to add regularization terms that penalize overly large triggers or scattered triggers. Authors in ~\cite{huang2019neuroninspect} show an improvement by also enhancing the NeuralCleanse optimization problem with additional regularization terms. Even though these methods improve certain aspects of NeuralCleanse, they include limited evaluation on complex localized triggers and novel triggers that are not applied in the conventional patch-based approach. For example, Instagram filters and smooth triggers are not patch-based and are applied in a different way. Note that the Clean-Label Attack~\cite{turner2018clean} and Hidden Backdoor Attack~\cite{saha2020hidden} are still patch-based trigger attacks. Hence, these defenses fail to detect triggers from different categories.  Moreover, they can incur a high computational cost that makes using the defense very laborious~\cite{shen2021backdoor}.

Offline defenses that focus on erasing backdoors can lead to degradation of the standard accuracy of the model and the removal of the backdoor. The current state-of-the-art in this line of work~\cite{li2021neural} removes the backdoor with up to
3\% degradation on the standard accuracy. In \cite{tran2018spectral, hayase2021spectre} the authors perform statistical analysis over the training data to identify Trojaned images and possibly remove them during training. ActivationClustering~\cite{chen2018detecting} showed that the poisoned training images can create a separate cluster from the clean images. Assuming that the defender has access to the training data is not realistic since the model can be trained by a third party. MNA~\cite{xu2019detecting} trains a meta classifier to detect Trojaned models. Recently, in ~\cite{zeng2021rethinking} the authors introduced smooth triggers that we evaluate on and have low frequencies, and show that MNA~\cite{xu2019detecting} has a low AUC when evaluated on smooth triggers.
% introduced a new method that can identify backdoors using the Discrete Cosine Transform components of a Trojaned image. Their main idea is that Trojaned images tend to exhibit high-frequency components compared to the clean images. They also 
% \penny{~\cite{kolouri2020universal, huster2021top}}
%~\cite{wang2019neural,chen2019deepinspect,NEURIPS2019_78211247,chou2020sentinet}
%analyze the network to identify whether it can be controlled by a Trojan trigger. However, these methods have high computational cost and limitations based on the number of labels~\cite{wang2019neural}, trigger's size or location in the image~\cite{chou2020sentinet} and often cannot detect complex triggers such as dynamic triggers~\cite{salem2020dynamic}. Other methods~\cite{gao2019strip,doan2019februus} perform run-time analysis to identify potentially Trojan triggers present in the inference-time images. 
\paragraph{Online Defenses.}
Online methods aim at detecting whether the model is Trojaned during inference.
These methods seem more promising for defending against different types of Trojan triggers since they have the advantage of encountering actual Trojaned images. In SentiNet~\cite{chou2020sentinet}, the authors propose a defense for localized triggers by examining the saliency maps in the input space. 
%Still, as mentioned above, the attacker can choose any trigger besides the traditional localized attack.
CleaNN~\cite{javaheripi2020cleann} leverages the observation that certain Trojaned images exhibit high frequencies in the frequency domain. However, the authors in~\cite{zeng2021rethinking} introduced the low-frequency trigger (mentioned as smooth trigger in our experiments) designed to break this assumption. In STRIP~\cite{gao2019strip}, the authors introduced a simple approach for distinguishing between Trojaned and clean images by superimposing the inference-time image with random clean images and checking the entropy of the resulting predictions. However, their approach relies on the fact that the two entropy distributions computed by Trojaned and clean images' labels can be separated by a threshold and its effectiveness is known to be very sensitive to this threshold.
%In our experiments, we find that the suggested default thresholds cannot distinguish clean and Trojaned images.
In our experiments, we find that the two entropy distributions cannot be cleanly separated and can have a significant overlap. We hypothesize that this poor performance is caused by their perturbation step which superimposes (adds) the input image with a random clean image. This simple addition often results in significantly out-of-distribution images and in some cases even cancels out the trigger. 
%Even by normalizing the images, the approach doesn't guarantee that the neural network will output the target label in Trojaned images or a range of different outputs in the case of clean images. 
% We also observe that the STRIP paper only considers triggers that do not overlap with the main part of the image or triggers for grayscale images with a black background where superimposing two images will not alter the actual trigger. 
% In Section~\ref{experiments}, we show that superimposing colored images can significantly alter the Trojan trigger.
Februus~\cite{doan2019februus}, uses attributions in the input space to identify Trojan triggers. As we show in this paper, relying on attributions at the input layer alone can't detect certain triggers. NEO~\cite{udeshi2019model} is an input filtering method that makes a strong assumption on the trigger type, assuming that triggers are patch-based and localized in order to identify their position and block them. Lastly, NNoculation~\cite{veldanda2020nnoculation} produces a second network from the potentially Trojaned network, by re-training to be robust to random perturbations. They make an interesting observation that this simple approach erases most of the backdoor's behavior (ASR drops to 2-8\%). Then, they identify disagreements between the two deployed models and eventually reverse-engineer the trigger using a Cycle-GAN. However, it is not clear if this approach can reverse-engineer non-localized triggers. Additionally, this approach requires 5\% of the inference images to be Trojaned to be effective.

\section{Discussion}
\label{discussion}
In this section, we discuss the possibility of an adaptive attack where the attacker would choose a trigger that, when added to input images, causes the attributions of the resulting Trojaned image to be similar to clean images' attributions, thereby avoiding detection by MISA. 
%One possible approach is to optimize the trigger to produce Trojaned attributions similar to clean attributions for all layers after a standard model is trained and then retrain the model with the resulted trigger. 
%However, we cannot force attributions to be similar for all layers and the function that computes attributions is not differentiable (Integrated Gradients is approximated by a discrete sum and cannot be computed exactly). 
Given that the function of computing attributions itself (e.g. Integrated Gradients is approximated by a discrete sum) is not differentiable, one possible approach is to use an additional loss term in the training objective to penalize high attributions on the Trojan pixels or features. 
%to penalize the attributions of the input in the trigger area during training using an additional regularization term in the loss function. 
In this case, the attacker must have control over the whole training process, as opposed to simply poisoning a small percentage of the training data as described in our threat model.
Even if we assume an attacker is able to implement such an attack,
this approach would require computing attributions for any intermediate model at each iteration during training, which will drastically slow down the training process. 
%even if we assume an attacker is able implement such an attack, our SVM is designed to classify most clean images as Trojaned before the extract-and-evaluate step as shown in Section~\ref{nostage3} and Table~\ref{tab:no_stage3}. 
%We also believe that the model will still produce anomalous attributions in another layer apart from the input layer of the neural network as evidenced by our experiments in Section~\ref{advantage}. 
Finally, as shown in this paper, misattribution is an inherent property of Trojaned images -- the trigger must be effective in a targeted attack regardless of what the clean part of the image is. 
Thus, such an attack is unlikely to produce a high attack success rate, and for the cases that it succeeds, would still be caught by the extract-and-evaluate stage of the MISA pipeline for a layer of the neural network. 
We leave a more comprehensive investigation of adaptive attacks to future work.
%and does not apply to clean features; hence, the extract-and-evaluate step would render an adaptive attack ineffective. Finally, including information of attributions during training is beyond the capability of the attacker according to our Threat Model, i.e., the attacker would need complete control over the training process as opposed to simply perturbing the training data.

\section{Conclusion}
\label{conclusion}
Our results demonstrate that we can successfully detect Trojaned instances at inference time without prior knowledge of the attack specifics by attributing the neural network's decision to input or intermediate-layer features. We observe that our method effectively detects different types of triggers, including recent ones that are not applied in the traditional patch-based approach, as explained in Section~\ref{preliminaries}. Our approach builds on the following two observations. First, attributions to a layer's features (input layer (pixels) or intermediate-layer features) of a Trojaned input are out-of-distribution from the clean features' attributions. Second, the target label persists when the high attributed Trojaned features are injected into the corresponding features from clean images.

\begin{acks}
This effort was supported by the Intelligence Advanced Research Projects Agency (IARPA) under the contract W911NF20C0038. 
The content of this paper does not necessarily reflect the position or the policy of the Government, and no official endorsement should be inferred.
\end{acks}

\bibliographystyle{ACM-Reference-Format}
\bibliography{./bib/paper.bib}

\appendix
\section{Appendix}
\begin{table*}[htb]
\caption{Model Architectures.}
\label{tab:architectures}
\centering
\begin{tabular}{c|l}
\hline 
    Dataset & NN Architecture \\
    \hline
    \multirow{4}{*}{MNIST} & Conv2D(32, (3,3)) + ReLU + \\
    & Conv2D(64, (3, 3)) + ReLU + \\
    & MaxPooling2D(2,2) + Dropout +\\
    & Dense(128) + ReLU + Dropout + Dense(10) + Softmax\\
    \hline
    \multirow{3}{*}{Fashion MNIST} & Conv2D(64, (12, 12)) + ReLU + MaxPooling2D(2, 2) + Dropout + \\ 
    & Conv2D(32, (8, 8)) + ReLU + MaxPooling2D(2, 2) + Dropout + \\
    & Dense(256) + ReLU + Dropout + Dense(10) + Softmax\\
    \hline
    \multirow{4}{*}{CIFAR10} & Conv2D(32, (3, 3)) + ReLU + BatchNorm + Conv2D(32, (3, 3)) + ReLU + BatchNorm + MaxPooling2D(2,2) + Dropout + \\
    & Conv2D(64, (3, 3)) + ReLU + BatchNorm + Conv2D(64, (3, 3)) + ReLU + BatchNorm +
    MaxPooling2D(2,2) + Dropout + \\
    & Conv2D(128, (3, 3)) + ReLU + BatchNorm + Conv2D(128, (3, 3)) + ReLU + BatchNorm +
    MaxPooling2D(2,2) + Dropout + \\
    & Dense(10) + Softmax \\
    \hline
    \multirow{4}{*}{GTSRB} & Conv2D(8, (5, 5)) + ReLU + BatchNorm + MaxPooling2D(2, 2) + \\
    & 2(Conv2D(16, (3, 3)) + ReLU + BatchNorm + MaxPooling2D(2, 2)) + \\
    & 2(Conv2D(32, (3, 3)) + ReLU + BatchNorm + MaxPooling2D(2, 2)) + \\
    & Dense(128) + ReLU + BatchNorm + Dropout + Dense(43) + Softmax\\
\hline
\end{tabular}
\end{table*}
\subsection{Neural Network Architectures}
This section shows the Neural Network architectures we used to train the Trojaned models for MNIST, Fashion MNIST, and GTSRB, as shown in Table~\ref{tab:architectures}.

\begin{figure}[!htp]
    \centering
    \includegraphics[width=0.97\linewidth]{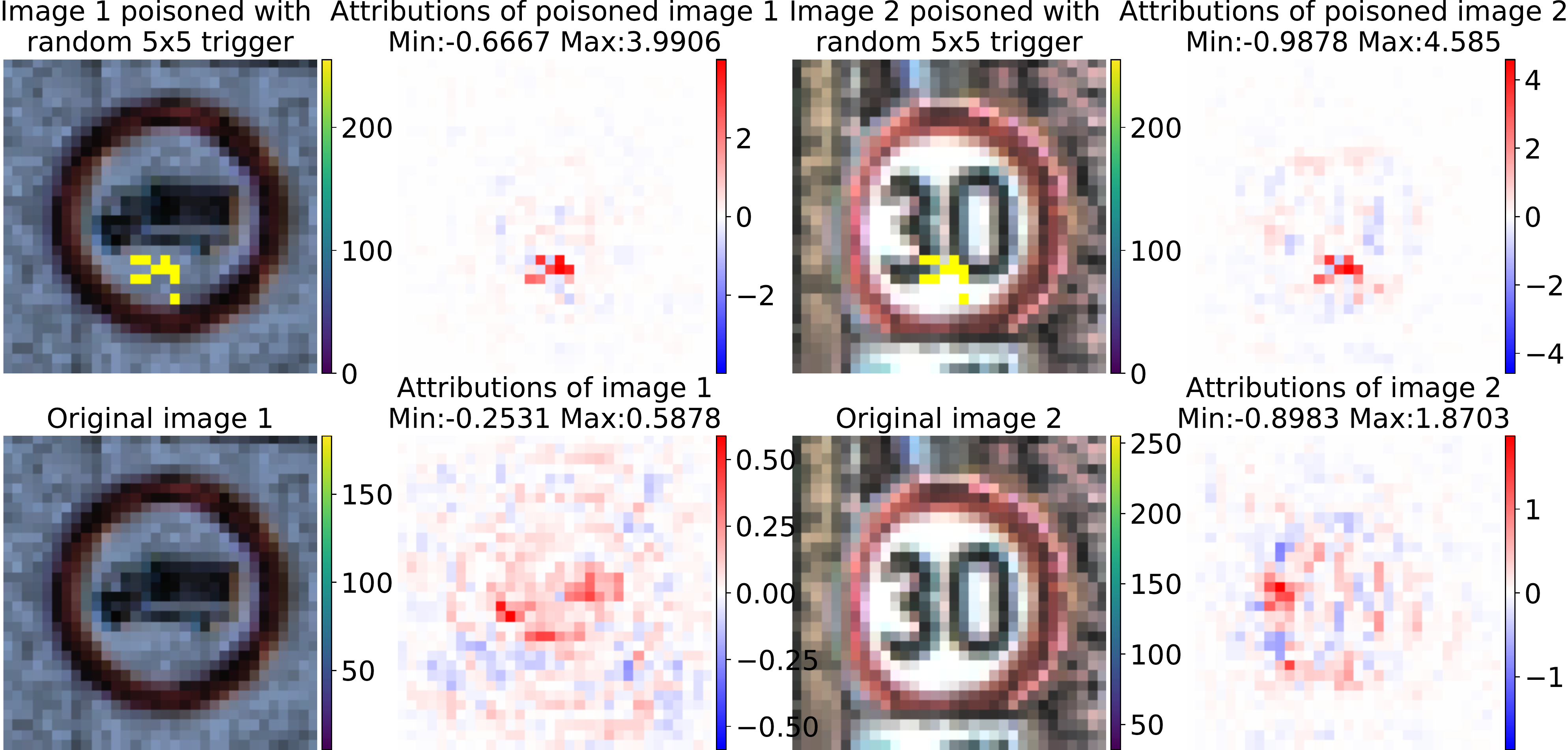}
    \caption{Example images poisoned with a random Trigger that is concentrated inside a 5x5 square. The top row shows the 2 poisoned images and their corresponding attribution maps (on the right of each image) derived from the image and the Trojaned model. The second row shows the same images without the trigger and their corresponding attribution maps derived from the image and the same Trojaned model.}
    \label{fig:images_with_random}
\end{figure}
\subsection{Triggers}
\subsubsection{Static Triggers}
Static triggers (applied in the patch-based approach) are placed in the same location every time we poison an image. We consider triggers that fully, partially, and barely obstructs/overlaps with the main part of the image. These locations include Top Middle (TM), Center Middle/Bottom Middle (M), and Bottom Right (BR) of an image. 

For grayscale images (MNIST, Fashion MNIST), we use white and gray trigger colors, while for colored images, we use yellow, purple, and white triggers with a black or blue background as well as randomly colored triggers. Fig.~\ref{fig:images_with_random} shows example images poisoned with a random trigger and their attribution map.
\begin{figure}[h]
    \centering
    \includegraphics[width=0.97\linewidth]{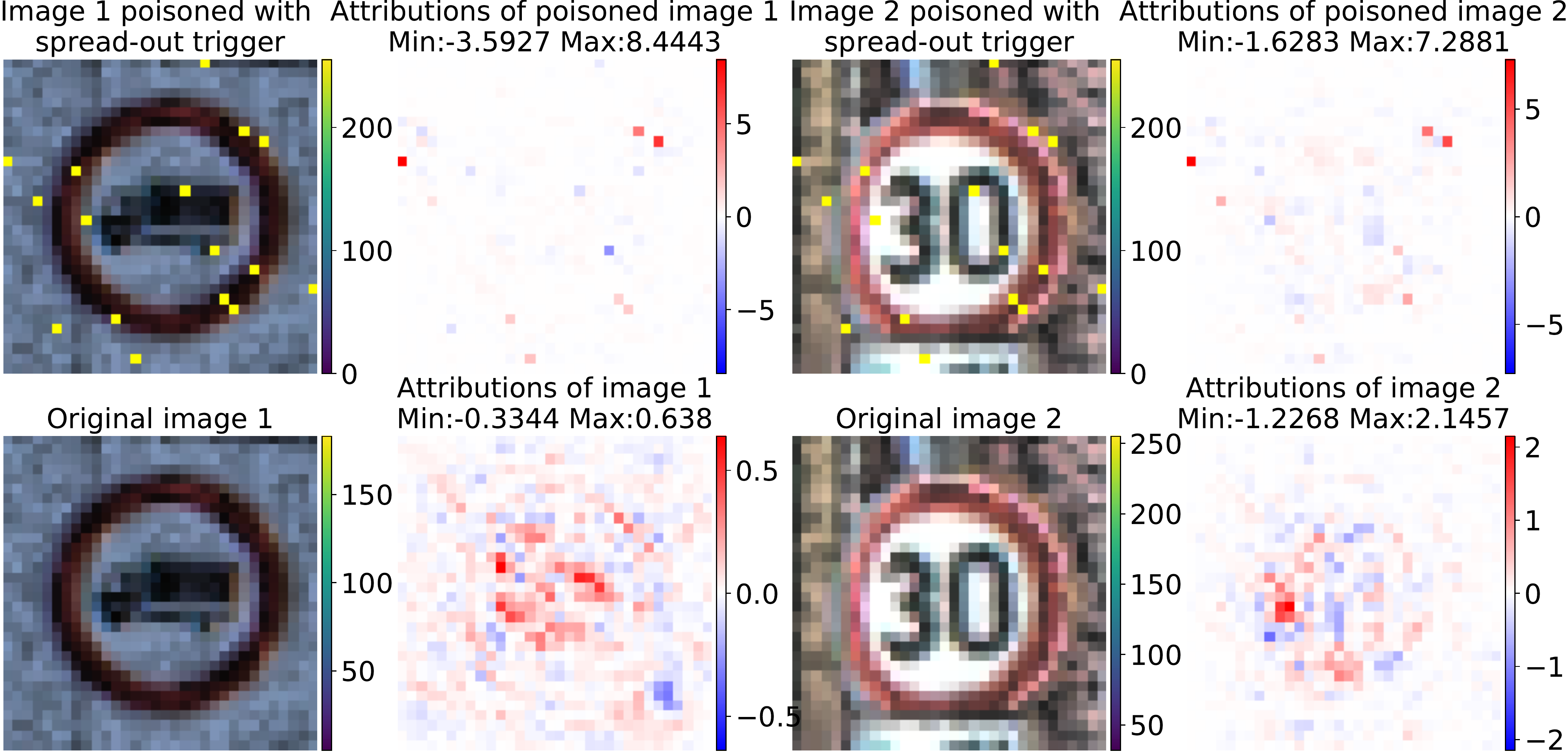}
    \caption{Example images poisoned with a spread-out Trigger of 16 pixels. The top row shows the 2 poisoned images and their corresponding attribution maps (on the right of each image) derived from the image and the Trojaned model. The second row shows the same images without the trigger and their corresponding attribution maps derived from the image and the same Trojaned model.}
    \label{fig:images_spread-out}
\end{figure}
\begin{figure}[h]
    \centering
    \includegraphics[width=0.97\linewidth]{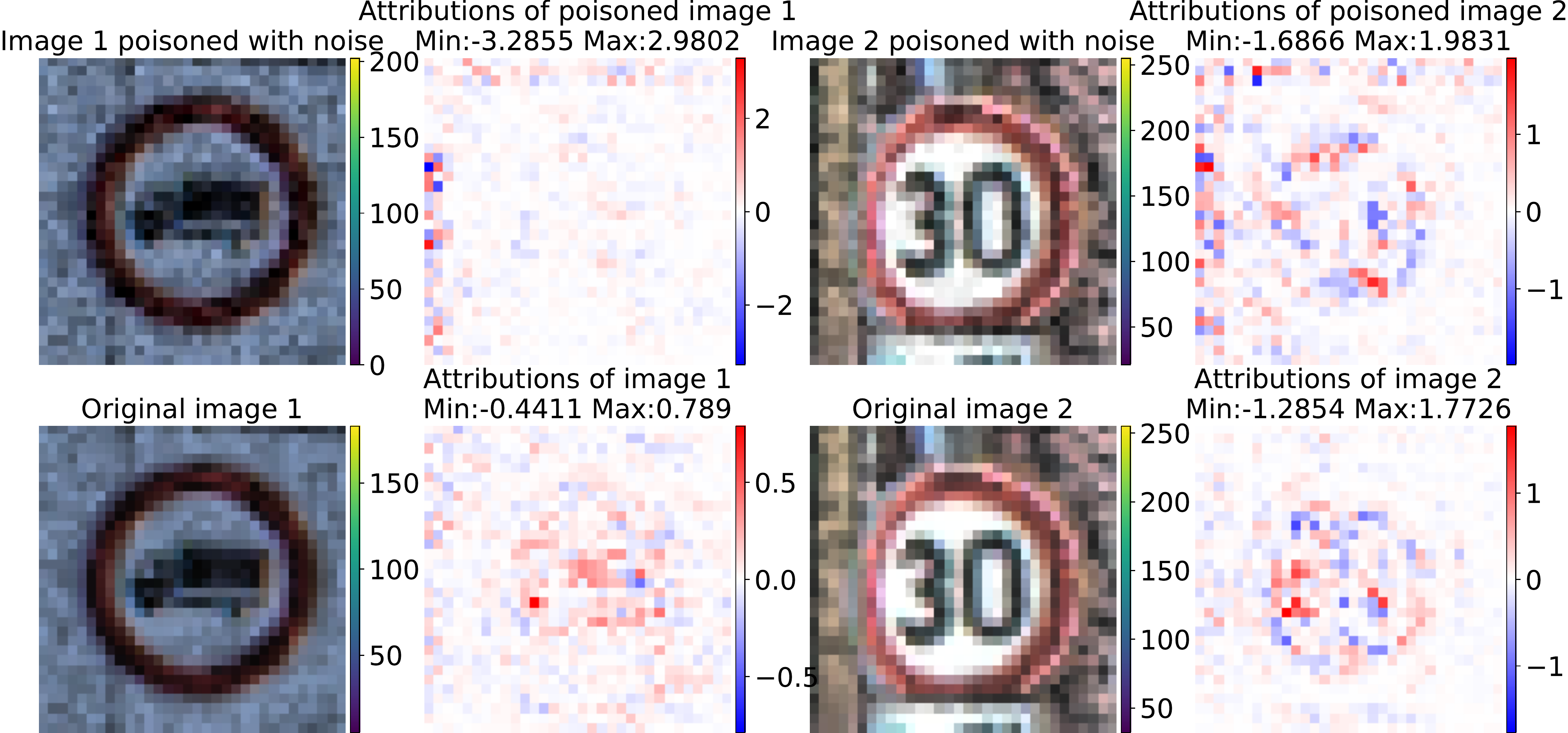}
    \caption{Example images poisoned with noise. The top row shows the 2 poisoned images and their corresponding attribution maps (on the right of each image) derived from the image and the Trojaned model. The second row shows the same images without the trigger and their corresponding attribution maps derived from the image and the same Trojaned model.}
    \label{fig:images_with_noise}
\end{figure}

For spread-out triggers we randomly choose $n$ pixels spread out in the image to change their color to white or yellow. In our experiments, $n$ is between 9 to 16. Fig.~\ref{fig:images_spread-out} shows example images poisoned with a spread-out trigger and their attribution map. We can see from Fig.~\ref{fig:images_spread-out} that the reverse-engineered trigger from high-attributed values will not be the actual trigger which explains the low TPR of our method. However, intermediate-layer attributions improve the TPR of our method as mentioned in the experimental section.

We refer to Instagram filters, smooth triggers and noise triggers as static as they don't require specifying a particular location. 
\begin{figure}[h]
    \centering
    \includegraphics[width=0.99\linewidth]{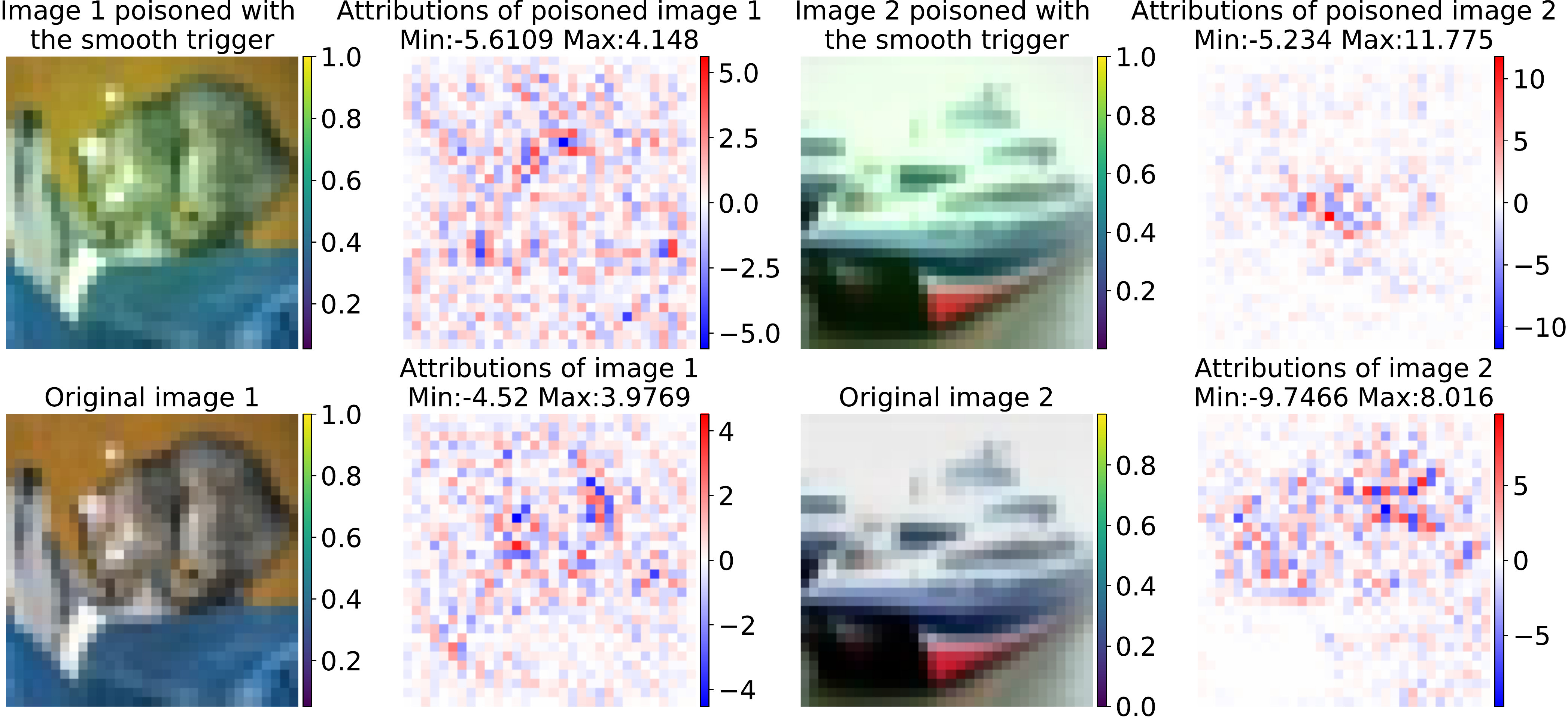}
    \caption{Example images poisoned with the smooth trigger. The top row shows the 2 poisoned images and their corresponding attribution maps (on the right of each image) derived from the image and the Trojaned model. The second row shows the same images without the trigger and their corresponding attribution maps derived from the image and the same Trojaned model.}
    \label{fig:smooth_triggers}
\end{figure}
\begin{figure}[h]
    \centering
    \includegraphics[width=0.99\linewidth]{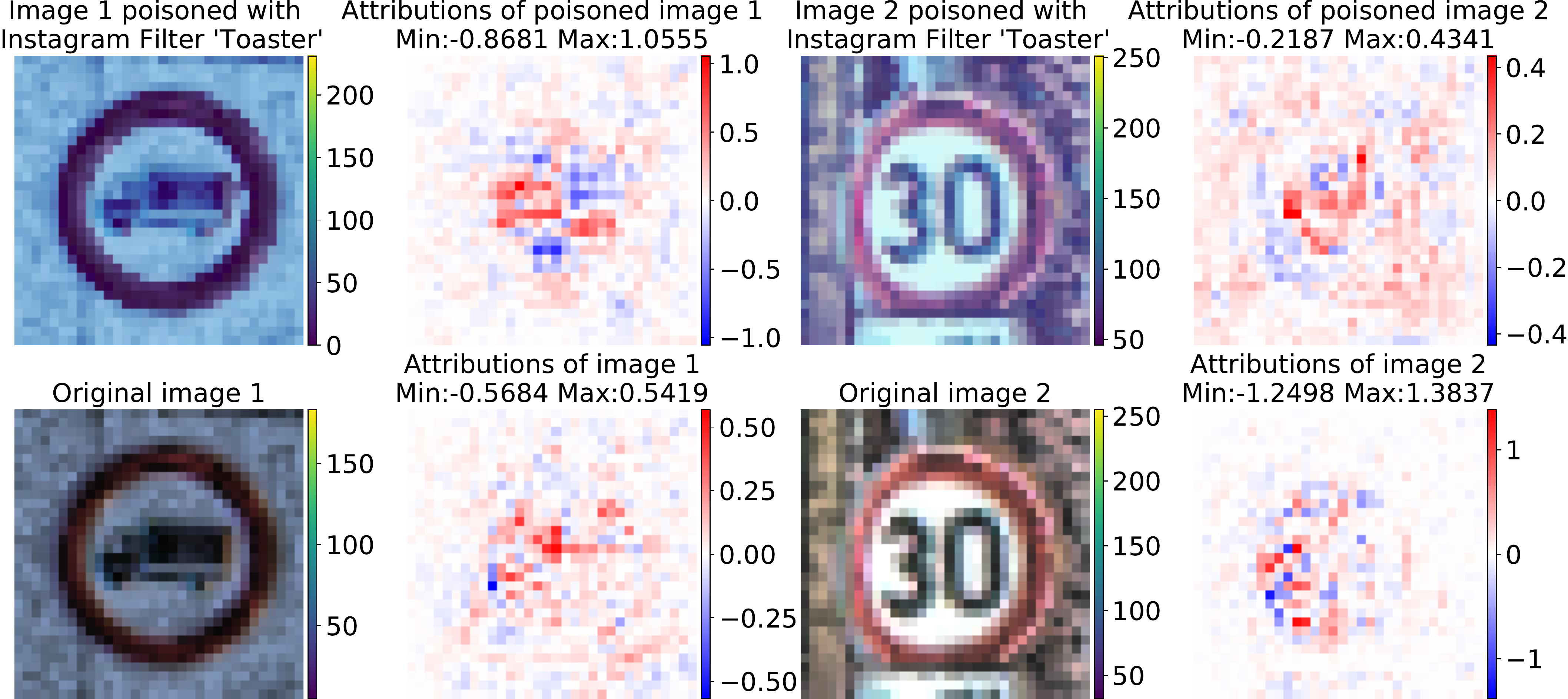}
    \caption{Example images poisoned with the Instagram filter Toaster. Example images poisoned with noise. The top row shows the 2 poisoned images and their corresponding attribution maps (on the right of each image) derived from the image and the Trojaned model. The second row shows the same images without the trigger and their corresponding attribution maps derived from the image and the same Trojaned model.}
    \label{fig:instagram_filters}
\end{figure}

Image-based triggers (smooth triggers and noise triggers) have the same size as the input. Instagram filters apply a transformation to the input. For noise triggers, we follow the approach of~\cite{chen2017targeted} to add random noise to the image: $\widetilde{\bm{x}} = \bm{x} + \bm{\delta}, \bm{\delta} \in [-20, 20]^{H\times W \times 3}$, where $\bm{\delta}$ is determined randomly. In Fig.~\ref{fig:images_with_noise} we provide 3 example images poisoned with noise and their corresponding attribution maps. For smooth triggers we used the trigger provided by the authors of~\cite{zeng2021rethinking} and also produced 10 more smooth triggers using their approach. Example images poisoned with smooth triggers along with their attribution map over the input layer is shown in Fig~\ref{fig:smooth_triggers}. As shown in the experimental section the attributions over the input layer don't reveal this type of trigger.

Finally, we used the following Instagram filters: Skyline, Toaster, and Walden.
Example images poisoned with Instagram Filters along with their attribution map over the input layer is shown in Fig~\ref{fig:instagram_filters}. As shown in the experimental section the attributions over the input layer don't reveal this type of trigger.

\subsubsection{Dynamic Triggers}
\begin{figure}[ht]
    \centering
    \includegraphics[width=0.97\linewidth]{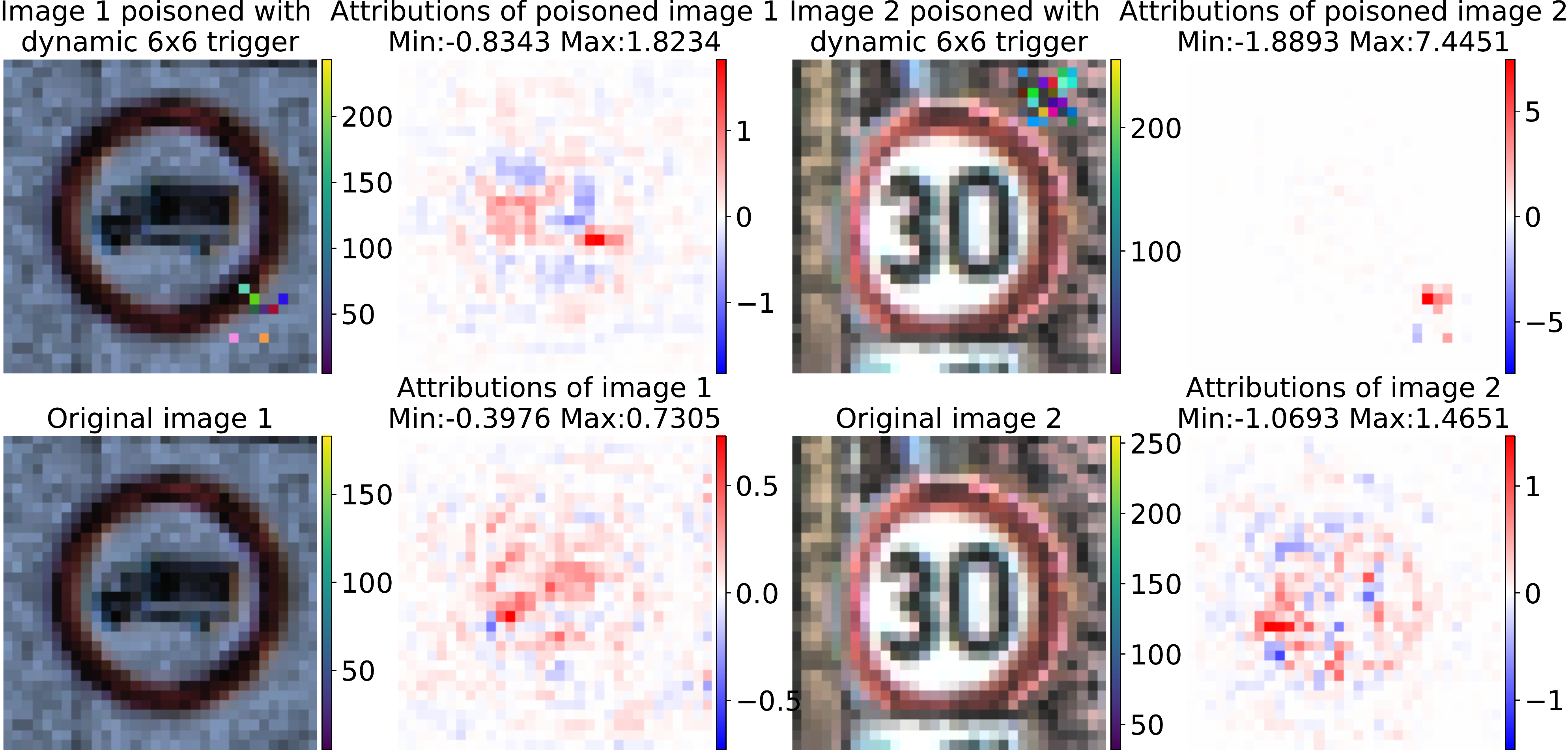}
    \caption{Example images poisoned with a dynamic trigger. The top row shows the 2 poisoned images and their corresponding attribution maps (on the right of each image) derived from the image and the Trojaned model. The second row shows the same images without the trigger and their corresponding attribution maps derived from the image and the same Trojaned model.}
    \label{fig:dynamic_triggers}
\end{figure}

Dynamic triggers can be generated by sampling a trigger from a set of triggers and a location from a predefined set of locations every time we poison an image~\cite{salem2020dynamic}. The set of triggers consists of triggers with random values for a given height and width generated by the TrojAI tool. The set of predefined locations consists of 9 locations scattered throughout the image, representing combinations of top, middle, bottom with left, center, right. We trained 12 models for each dataset, with dynamic triggers of shape 3x3, 5x5, 6x6, and 8x8. The set of triggers includes 10 triggers that are created by choosing a random assignment of values between 0 and 255. Example dynamic triggers with the corresponding clean images and their attributions are shown in Fig~\ref{fig:dynamic_triggers}.

%We provide the average Detection Accuracies for MNIST, Fashion MNIST, and GTSRB in Table~\ref{tab:results_dynamic_trigger}. 

In Table~\ref{tab:svm_details}, we give the number of attribution maps used for training the SVMs. For example, for each MNIST Trojaned model, we trained one SVM on the 8000 clean attributions that were derived from the model and the corresponding 8000 clean images of the evaluation set. 
\begin{table}[htb]
\caption{Number of attribution maps (derived from clean images) involved in the training of each SVM.}
\label{tab:svm_details}
\centering
\begin{tabular}{c|c}
\hline 
    Dataset & \# SVM training instances \\
    \hline
    MNIST & 8000 \\
    \hline
    Fashion MNIST & 8000 \\
    \hline
    CIFAR10 & 8000 \\
    \hline
    GTSRB & 10104 \\
\hline
\end{tabular}
\end{table}
\subsection{Examples supporting Experimental Results}
\begin{figure}[ht]
    \centering
    \includegraphics[width=0.97\linewidth]{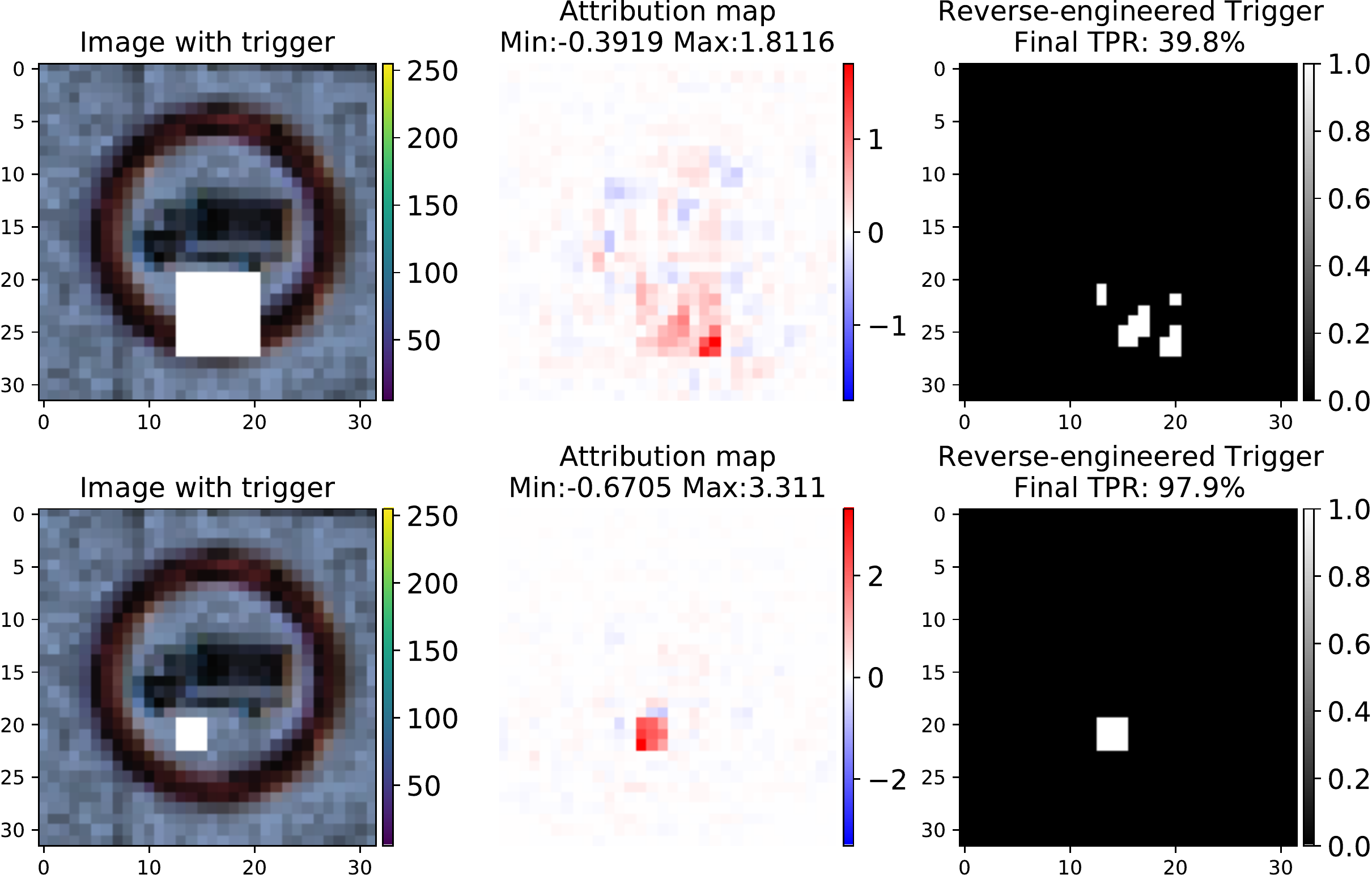}
    \caption{Comparison of reverse-engineered triggers for large and small square triggers.
    Top row: image with 8x8 white square trigger, attribution map, and reverse-engineered trigger. Second row: image with 3x3 white square trigger, attribution map, and reverse-engineered trigger.}
    \label{fig:large_trigger_issues}
\end{figure}

In Fig.~\ref{fig:large_trigger_issues} we compare the attributions from large and small triggers. As mentioned in the experimental section larger triggers are not always reverse-engineered correctly.

\begin{figure}[ht]
    \centering
    \includegraphics[width=0.97\linewidth]{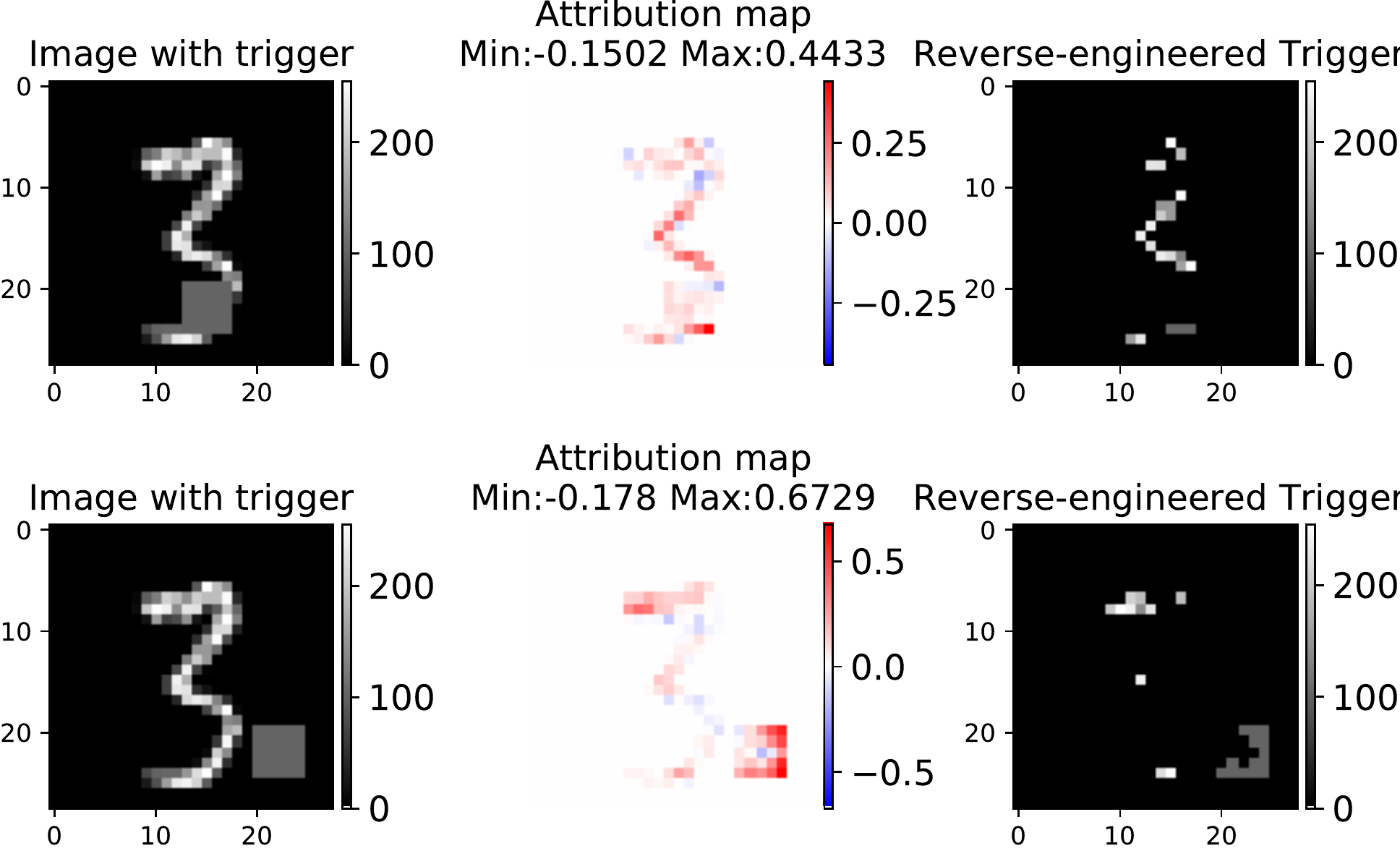}
    \caption{Comparison of reverse-engineered triggers for different positions of the same trigger. 
    Top row: image with trigger in the center. Second row: Image with trigger in the bottom right.
    }
    \label{fig:mnist_trigger_location}
\end{figure}

In Fig.~\ref{fig:mnist_trigger_location} we show how the location of grayscale triggers can affect the attribution and the ability of the method to recover a good portion of the trigger. As mentioned in the experimental section this happens mostly due to the use of a black baseline. 

% Fig.~\ref{fig:gtsrb_grad_cam_comparison} presents Grad-CAM and DeepSHAP attributions for the same image. Using Grad-CAM attributions we extract clean features as triggers that don't include the main part of the Trojaned trigger.
% \begin{figure}[ht]
%     \centering
%     \includegraphics[width=0.70\linewidth]{./figures/gtsrb_grad_cam_comparison.pdf}
%     \caption{Comparison between Grad-CAM and DeepSHAP attribution maps for the same image.}
%     \label{fig:gtsrb_grad_cam_comparison}
% \end{figure}

\begin{figure}[ht]
    \centering
    \includegraphics[width=0.97\linewidth]{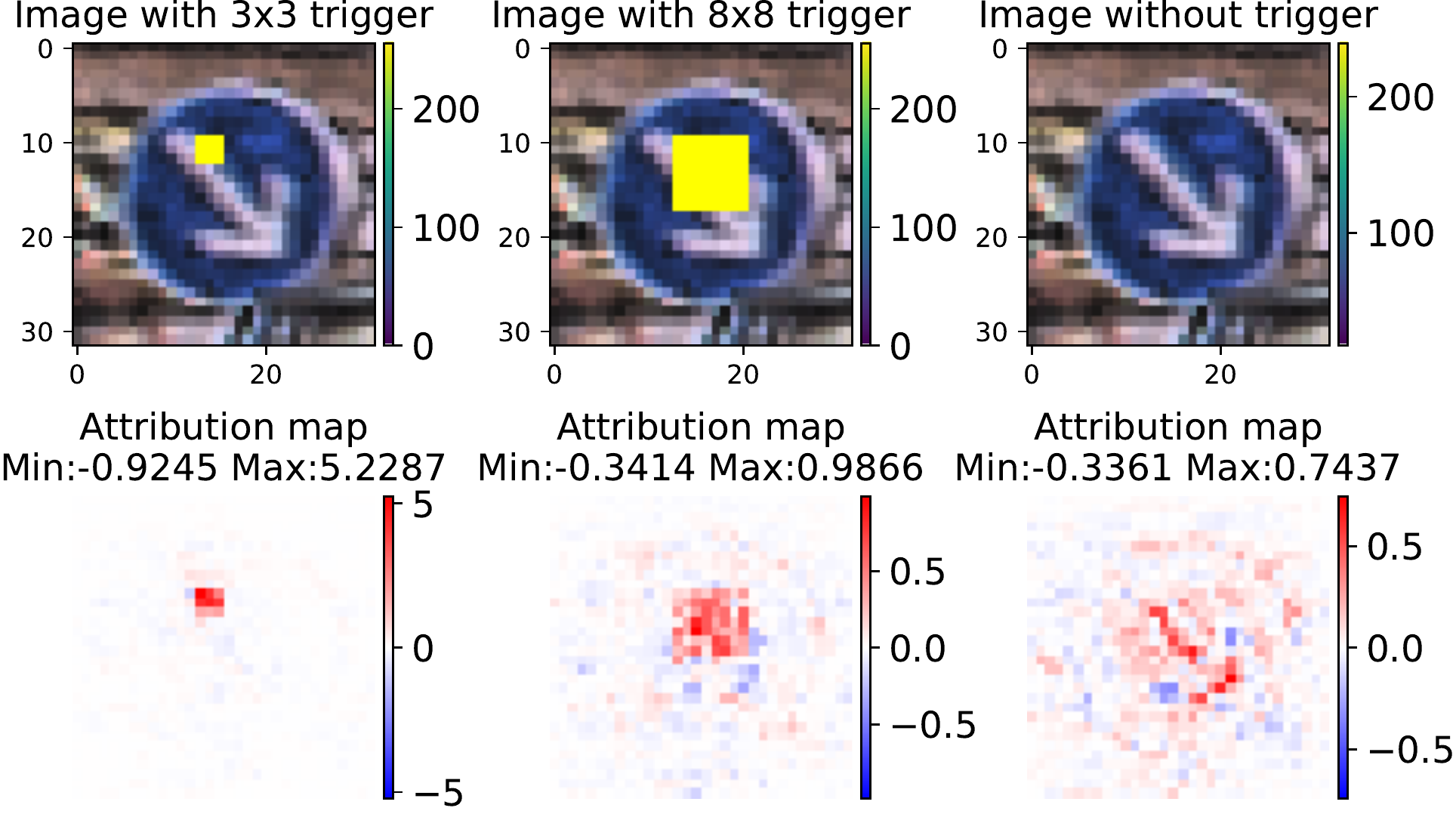}
    \caption{Comparison of attribution maps between clean and Trojaned images for different sizes of a trigger added in the center.
    and range of values for an image with a small trigger, an image with a large trigger, and a clean image. Top row: image with and without the triggers. Second row: Corresponding attribution maps.}
    \label{fig:gtsrb_diff_clean_trojan}
\end{figure}

\subsection{Percentage of Poisoning during Training}
\begin{figure}[H]
    \centering
    \includegraphics[width=0.99\linewidth]{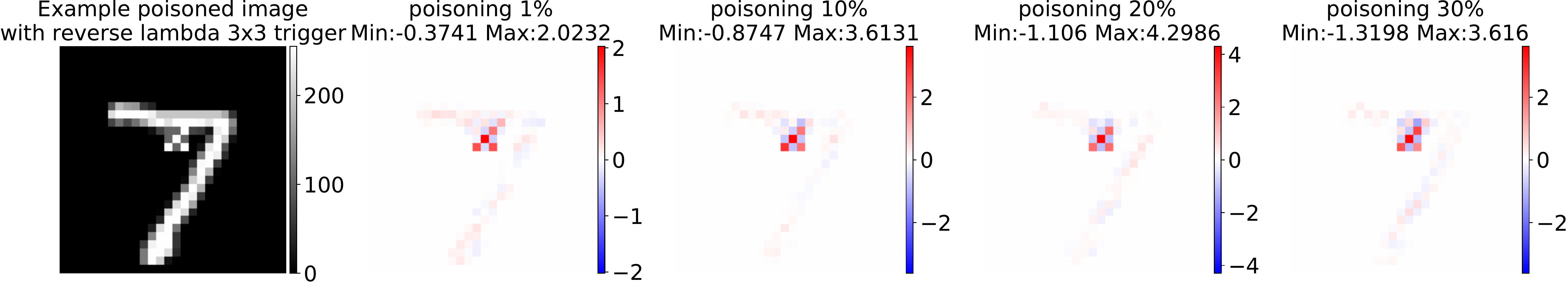}
    \caption{Attribution values for the reverse lambda Trigger of size 3x3 across models with increasing percentage of poisoning during training.}
    \label{fig:mnist_lambda_3x3}
\end{figure}

\begin{figure}[H]
    \centering
    \includegraphics[width=0.99\linewidth]{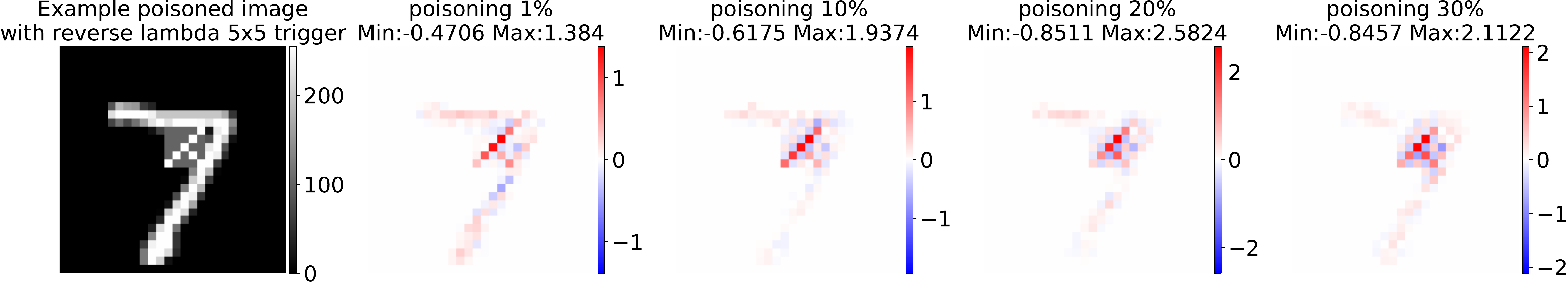}
    \caption{Attribution values for the reverse lambda Trigger of size 5x5 across models with increasing percentage of poisoning during training.}
    \label{fig:mnist_lambda_5x5}
\end{figure}

\begin{figure}[H]
    \centering
    \includegraphics[width=0.99\linewidth]{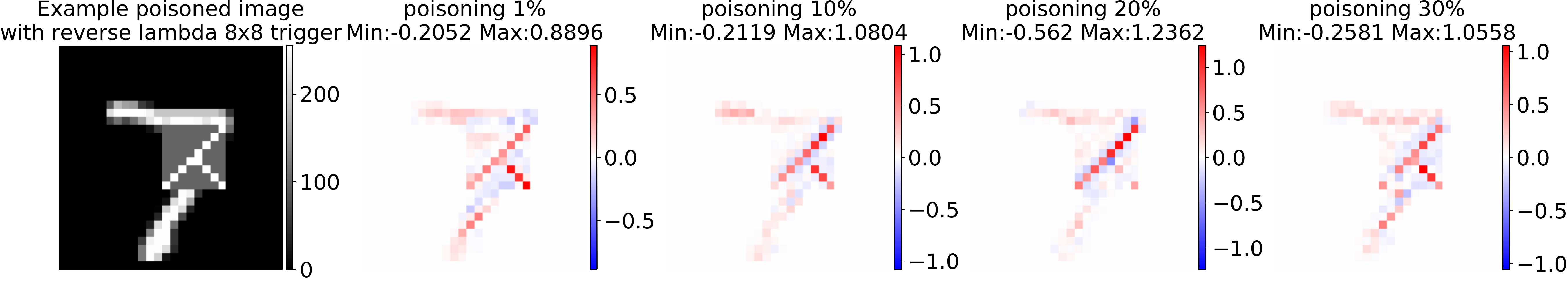}
    \caption{Attribution values for the reverse lambda Trigger of size 8x8 across models with increasing percentage of poisoning during training.}
    \label{fig:mnist_lambda_8x8}
\end{figure}

\begin{figure}[H]
    \centering
    \includegraphics[width=0.97\linewidth]{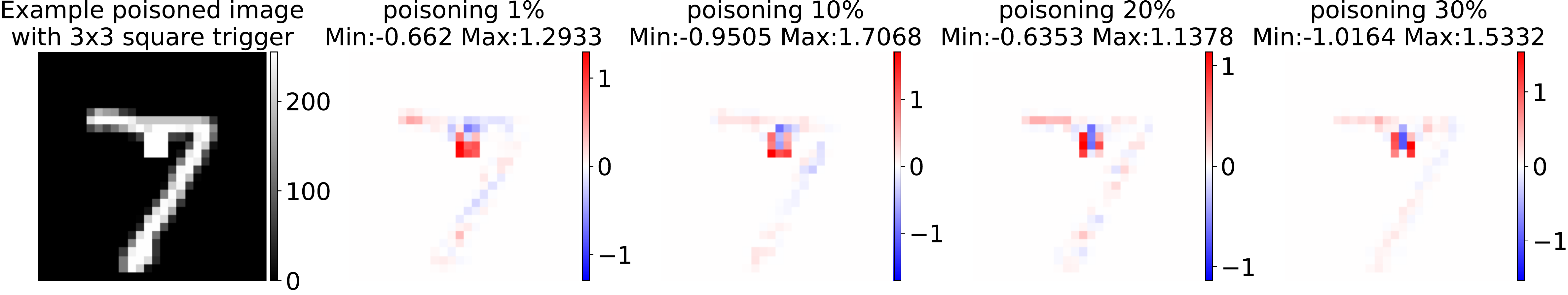}
    \caption{Attribution values for the square Trigger of size 3x3 across models with increasing percentage of poisoning during training.}
    \label{fig:mnist_rectangle_3x3}
\end{figure}

\begin{figure}[H]
    \centering
    \includegraphics[width=0.95\linewidth]{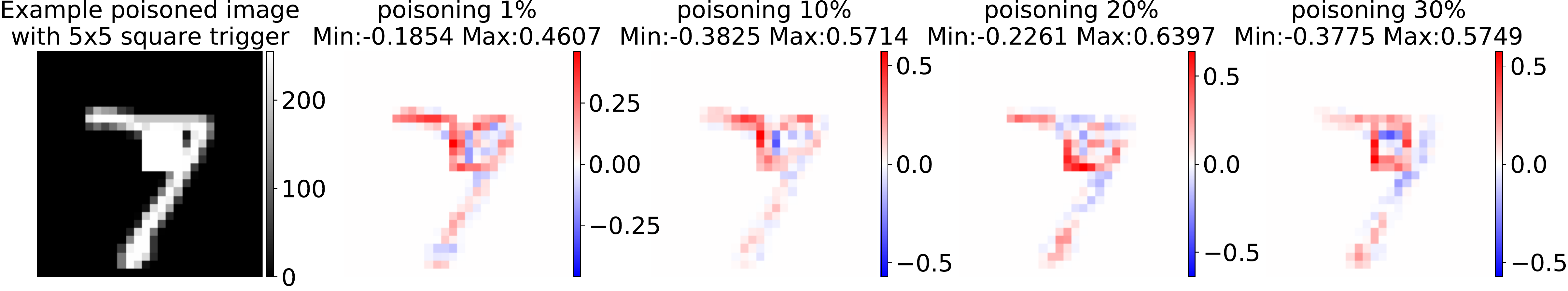}
    \caption{Attribution values for the square Trigger of size 5x5 across models with increasing percentage of poisoning during training.}
    \label{fig:mnist_rectangle_5x5}
\end{figure}

\begin{figure}[H]
    \centering
    \includegraphics[width=0.95\linewidth]{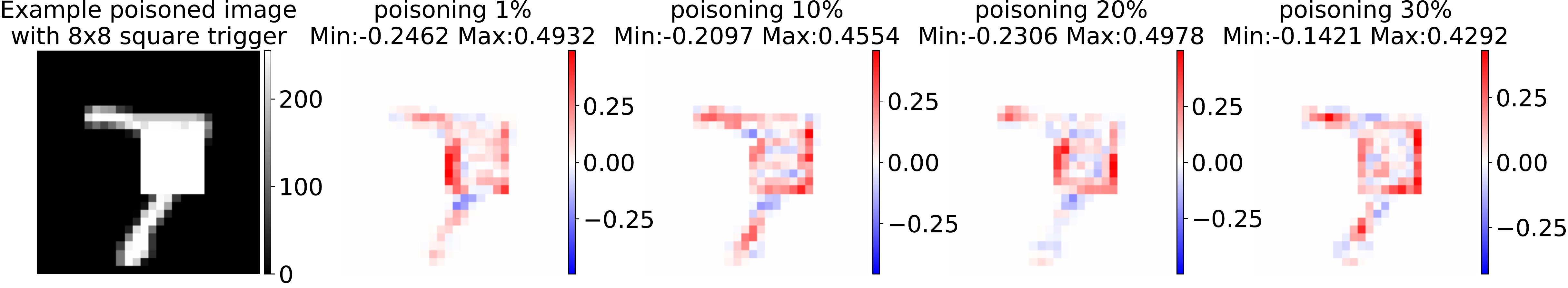}
    \caption{Attribution values for the square Trigger of size 8x8 across models with increasing percentage of poisoning during training.}
    \label{fig:mnist_rectangle_8x8}
\end{figure}

In this paper, we poison the minimum number of training instances required to obtain a Trojan model. We provide the percentage of training instances that are poisoned in Table~\ref{tab:percentage_of_poisoning}. We observe that increasing this percentage can lead to higher attribution values for the Trojan trigger in certain cases, as shown in Figures~\ref{fig:mnist_lambda_3x3},~\ref{fig:mnist_lambda_5x5},~\ref{fig:mnist_lambda_8x8},~\ref{fig:mnist_rectangle_3x3},~\ref{fig:mnist_rectangle_5x5}, and ~\ref{fig:mnist_rectangle_8x8}. 

\begin{table}[htb]
\caption{Percentage of training instances poisoned during training of the NN models.}
\label{tab:percentage_of_poisoning}
\centering
\begin{tabular}{c|c|c}
\hline 
    Dataset & Trigger Type & Poisoning \\
    \hline
    \multirow{3}{*}{MNIST} & Static (except Noise) & 1\% \\
     & Noise & 20\% \\
     & Dynamic & 10\% \\
    \hline
    \multirow{3}{*}{Fashion MNIST} & Static  (except Noise) & 1\% \\
     & Noise & 20\% \\
     & Dynamic & 10\% \\
    \hline
    \multirow{5}{*}{GTSRB} & Static (except Noise) & 10\% \\
     & Noise & 20\% \\
     & Dynamic & 10\% \\
\hline
\end{tabular}
\end{table}


\end{document}